\documentclass[12pt,letterpaper]{article}
\usepackage[utf8]{inputenc}

\usepackage{color}
\definecolor{darkblue}{rgb}{0.1,0.1,.7}
\usepackage[colorlinks, linkcolor=darkblue, citecolor=darkblue, urlcolor=darkblue, linktocpage]{hyperref}
\usepackage[]{amsmath}
\usepackage[]{graphicx}
\usepackage[]{latexsym}
\usepackage[utf8]{inputenc}
\usepackage{slashed,graphicx,color,amsmath,amssymb}
\usepackage[mathscr]{eucal}
\usepackage{mathrsfs}
\usepackage[margin=0pt,font=small,labelfont=bf]{caption}
\usepackage{amscd}
\usepackage{bm}
\usepackage{xcolor}
\usepackage{mathtools}
\usepackage{upgreek}
\usepackage[square, comma, sort&compress,numbers]{natbib}
\usepackage[all,cmtip]{xy}
\usepackage[margin=1.59in]{geometry}
\usepackage{cleveref}
\usepackage[color=cyan!30!white,linecolor=red,textsize=footnotesize]{todonotes}
\usepackage{soul}
\numberwithin{equation}{section}
\hypersetup{
	pdfstartview={FitH},    
	pdftitle={Virasoro blocks and quasimodular forms},    
	pdfauthor={Das, Datta, Raman},     
	colorlinks=true,       
	linkcolor=blue,          
	citecolor=blue!59!black! ,        
	filecolor=magenta,      
	urlcolor=blue           
}

\newcommand{\tr}{\mathrm{Tr}\,}

\def\btau{{\bar{\tau}}}

\def\DD#1{{\mathscr{D}}^{(#1)}}

\def\bq{{\bar{q}}}

\def\bh{{\bar{h}}}

\def\bz{{\bar{z}}}

\def\nn{\nonumber}

\def\l1{{{1-loop}}}

\def\bz{{\bar{z}}}

\def\im{{{Im}}}

\def\n1{\Bigg|_{n=1}}

\def\n{{(n)}}
\def\tr{{Tr}}

\def\tr{\text{Tr}}
\def\th{\theta}

\def\bL{\bar{L}}

\def\bq{\bar{q}}

\def\cL{\mathcal{L}}

\def\bh{{\bar{h}}}
\def\bz{{\bar{z}}}

\def\th{\tilde{h}}

\def\beq{\begin{equation}}
\def\eeq{\end{equation}}

\def\bea{\begin{eqnarray}}
\def\eea{\end{eqnarray}}
\def\nn{\nonumber}

\def\l1{{\text{1-loop}}}

\def\bz{{\bar{z}}}

\def\im{{\text{Im}}}

\def\n1{\Bigg|_{n=1}}

\def\n{{(n)}}
\def\tr{\text{Tr}}

\def\bra#1{\left\langle #1 \right\rvert}

\def\ket#1{\left\lvert #1 \right\rangle}

\def\vev#1{\left\langle{#1}\right\rangle}

\def\cH{\mathcal{H}}

\def\bz{\bar{z}}

\def\cF{\mathcal{F}}

\def\be{\begin{equation}}
\def\ee{\end{equation}}
\def\bal{\begin{array}{l}}
\def\ba#1{\begin{array}{#1}}  
	\def\ea{\end{array}}
\def\bea{\begin{eqnarray}}
\def\eea{\end{eqnarray}}
\def\beas{\begin{eqnarray*}}
	\def\eeas{\end{eqnarray*}}

\def\nn{\\\nonumber}

\def\vev#1{\left\langle #1 \right\rangle}

\def\ceil#1{{\lceil #1 \rceil}}

\def\nn{\nonumber}
\def\bit{\begin{item}}
	\def\eit{\end{item}}
\def\benu{\begin{enumerate}}
	\def\eenu{\end{enumerate}}
\def\tr{{\rm tr}}

\def\bz{\bar{z}}

\def\cO{{\mathcal O}}

 \makeatletter
 \g@addto@macro\bfseries{\boldmath}
 \makeatother

\makeatletter
\if@todonotes@disabled

\else

\fi
\makeatother

\def\CO{{\mathcal{O}}}

\def\cV{\mathcal{V}}

\begin{document}

\definecolor{tinge}{RGB}{255, 244, 195}
\sethlcolor{tinge}
\setstcolor{red}

\definecolor{rust}{rgb}{0.5,0.1,0.1}
\def\DD#1{{\color{rust}{#1}}}

\vspace*{-1.2cm} \thispagestyle{empty}
\begin{flushright}
\texttt{CERN-TH-2020-119}
\end{flushright}
\vspace{.6cm} {\Large
\begin{center}
{\LARGE \bf  Virasoro blocks and quasimodular forms}
\end{center}}
\vspace{.2in}
\begin{center}
{Diptarka Das$^2$, Shouvik Datta$^{4}$ and Madhusudhan Raman$^6$}
\\
\vspace{.3in}
\small{
$^2$  \textit{Department of Physics, Indian Institute of Technology - Kanpur,\\
	Kanpur 208016, India.}
\\ \vspace{.1cm}
\vspace{.2cm}
$^4$ \textit{Department of Theoretical Physics, CERN,\\
	1 Esplanade des Particules, Geneva 23, CH-1211, Switzerland.}
}\\ \vspace{.1cm}
\vspace{.2cm}
$^6$ \textit{Department of Theoretical Physics, Tata Institute of Fundamental Research,\\
	Homi Bhabha Road, Navy Nagar, Colaba, Mumbai 400 005, India.
	 }
\\ \vspace{.1cm}
\vspace{.25in}
\begingroup\ttfamily\small
didas@iitk.ac.in,~sdatta@cern.ch,~madhur@theory.tifr.res.in\par
\endgroup

\end{center}

\vspace{.5in}

\begin{abstract}
\normalsize{We analyse Virasoro conformal blocks in the regime of heavy intermediate exchange $ (h_p \rightarrow \infty) $. For the $1$-point block on the torus and the $4$-point block on the sphere, we show that each order in the large-$ h_p $ expansion can be written in closed form as polynomials in the Eisenstein series. The appearance of this structure is explained using the fusion kernel and, more markedly, by invoking the modular anomaly equations via the $ 2 $d/$ 4 $d correspondence. We observe that the existence of these constraints allows us to develop a faster algorithm to recursively construct the blocks in this regime. We then apply our results to find corrections to averaged heavy-heavy-light OPE coefficients.}
\end{abstract}

\vskip 1cm \hspace{0.7cm}

\newpage

\setcounter{page}{1}

\noindent\rule{\textwidth}{.1pt}\vspace{-1.2cm}
\begingroup
\hypersetup{linkcolor=black}
\tableofcontents
\endgroup
\noindent\rule{\textwidth}{.2pt}

\section{Introduction}

Correlation functions in conformal field theories are built out of fundamental objects known as conformal blocks. These objects are fixed perturbatively by conformal symmetries and play a key role in various lines of research, including the AdS/CFT correspondence and the conformal bootstrap. In two spacetime dimensions, owing to an infinite-dimensional Virasoro symmetry, the conformal block has a rich structure. A closed form for the blocks, however, still eludes us except in a handful of special cases. 

For example, at large-$c$ (with conformal dimensions also scaling with $c$) the Virasoro blocks exponentiate \cite{zamu1,dattaper}, and in this case a closed form for the blocks can be found,  by using the monodromy method or the oscillator formalism \cite{zamolodchikov1986two}.\footnote{For contrast, when conformal dimensions are held fixed, the leading answer for blocks on the sphere in the large-$c$ limit is given by the global block, for which the closed form is known in terms of hypergeometric functions.} The monodromy method has also been used to determine blocks for heavy-light external dimensions \cite{fkw}, as well as to determine the blocks for asymptotically heavy intermediate dimensions \cite{harlow}.

In this paper, we study Virasoro conformal blocks in the regime of heavy intermediate exchange, i.e.~we study the blocks as an expansion in inverse powers of the conformal dimension of the exchanged operator. This study is similar in spirit to other asymptotic analyses in physics, where various simplifications and interesting features arise when quantities of interest are expanded in a large parameter, for example the rank of a gauge group. 

Our work draws on a number of themes, each of which we now discuss.

\subsubsection*{Zamolodchikov recursion}

An efficient means of computing the 4-point Virasoro block on the plane is via the recursion relations discovered by Zamolodchikov \cite{zamu1, zamu2}. This recursion is based on the structure of poles and residues of the block arising due to the presence of degenerate representations. The block can then be written as a sum over appropriately weighted poles of either the exchanged conformal dimension $h_p$, or the central charge $c$. In either case, the terms in the sum can be recursively evaluated, allowing for a {\em perturbative} determination of the block as an expansion in the cross-ratio $z$, or in the elliptic nome $q$ associated to the pillow coordinates. Similar strategies have been used to derive recursive representations of Virasoro blocks on the torus \cite{Hadasz:2009db} and on higher genus Riemann surfaces \cite{xy}. 

In both cases, however, the full non-perturbative (in either $z$ or in $q$) answer for the block is beyond reach. It would therefore be of interest to find further constraints satisfied by conformal blocks that, when combined with Zamolodchikov recursion, can be used to determine the block non-perturbatively (at least in principle). This brings us to the additional asset of modularity, to which we now turn.

\subsubsection*{Modularity}

It is well-known that conformal correlators enjoy modular properties. In the context of torus $ 1 $- and $ 2 $-point functions, this property has been used to find asymptotic formulae for OPE coefficients, pioneered by \cite{Km}, and later adapted to various other cases \cite{modcharged, modoff, modoff2, modoff3}. Additionally, since crossing symmetry of the full $ 4 $-point sphere correlator can be expressed as a modular property, asymptotic constraints can also be obtained from bootstrapping the high ``temperature'' result \cite{modpillow}. 

Conformal correlators are built out of conformal blocks weighted by OPE coefficients. On either side of the crossing equation, blocks in dual channels (or dual tori) appear.  A remarkable fact about two-dimensional CFTs is the existence of integral kernels which relate S-dual blocks \cite{Ponsot:1999uf, Ponsot:2000mt}. These have been used recently to bootstrap the CFT data \cite{collier-mod, brehmdas, maloney-mod}. 
It seems unlikely, however, that the Virasoro blocks themselves (on the torus, or on the sphere in the elliptic representation) will have any definite modular properties.\footnote{See, however, \cite{Cheng:2020srs} for a recent development.} If such a property exists in general, even partially, one might hope that a closed form expression for these blocks is possible. In this work we demonstrate that  
\begin{itemize}
	\item when the Virasoro blocks are expanded in a specific linear combination of the intermediate conformal dimension and the central charge, the coefficients of the  expansion can be resummed into quasimodular forms of PSL$ (2,\mathbb{Z}) $. 
	\item Further, these coefficients are constrained to satisfy a ``modular anomaly equation'' that one can use to recursively determine higher orders in this expansion, with minimal input from Zamolodchikov recursion.
	\item Finally, from this expansion one can read off the coefficients of the large-$ h_p $ expansion straightforwardly.
\end{itemize}

The closed form for the $ 4 $-point block on the sphere at leading order in large-$h_p$ was recently obtained using Zamolodchikov recursion in \cite{Cardona:2020cfy}. This was done by computing the first few orders in the $ q $-series explicitly and noting that it can be resummed into the quasimodular weight-$ 2 $ Eisenstein series, $E_2(\tau)$. Indeed, a similar result was originally established for both the four-point block on the sphere and the one-point block on the torus by \cite{KashaniPoor:2012wb,Kashani-Poor:2013oza,Kashani-Poor:2014mua}. We find that in a large-$ h_p $ expansion of the block, the coefficient of $ h_p^{-n} $ can be written as a linear combination of all possible quasi-modular forms of weights $2n$ and lower. As a result, these coefficients do not have definite modular weight. We shall show explicitly that a suitable reorganization of the large-$h_p$ expansion makes the modular features more manifest. Furthermore, the closed form expressions in terms of the Eisenstein series allow us to specify the block on the entire unit disk in the $ q $-plane.

The fact that the coefficients of the large-$ h_p $ expansion are constrained to satisfy a modular anomaly equation is explained by appealing to the 2d/4d correspondence \cite{agt}. This is briefly discussed below. 

\subsubsection*{Gauge theories, the $2$d/$4$d correspondence, and a synthesis}

Much effort has been directed towards understanding instanton effects in $ \mathcal{N} = 2 $ supersymmetric gauge theories. Notably, techniques have been developed to localise path integrals onto instanton moduli spaces, and further onto sets of isolated points, thereby allowing for their explicit evaluation, see \cite{Pestun:2016zxk} for an expansive review. These computations were made possible by the introduction of the $ \Omega $-background --- a specific supergravity background parametrised by $ (\epsilon_1,\epsilon_{2}) $ and with non-trivial graviphoton field strength --- which has the effect of regularising the volume of spacetime. Against this background, one can compute the deformed instanton partition function $ Z_{\text{inst.}}\left(\epsilon_{1}, \epsilon_{2}\right) $, from which the prepotential of the undeformed gauge theory is given by
		\begin{equation}\label{key}
		F_{\text {inst.}}=-\lim _{\epsilon_{1}, \epsilon_{2} \rightarrow 0} \epsilon_{1} \epsilon_{2} \log Z_{\text {inst.}}\left(\epsilon_{1}, \epsilon_{2}\right) \ .
		\end{equation}
		It has been known for some time now that the instanton prepotential $ F_{\text{inst.}} $ in a semiclassical expansion (for large values of the Coulomb moduli) can be resummed into quasi-modular forms of the relevant S-duality group \cite{Minahan:1997if,Billo:2015pjb,Ashok:2016oyh}. 
		
		Perhaps more interestingly, one can consider a \emph{deformed} prepotential
		\begin{equation}\label{key}
		F_{\text {inst.}}\left(\epsilon_{1}, \epsilon_{2}\right)=-\epsilon_{1} \epsilon_{2} \log Z_{\text {inst.}}\left(\epsilon_{1}, \epsilon_{2}\right) = \sum_{n, g=0}^{\infty} F^{(n, g)}\left(\epsilon_{1}+\epsilon_{2}\right)^{2 n}\left(\epsilon_{1} \epsilon_{2}\right)^{g} \ ,
		\end{equation}
		where the $ F^{(n,g)} $ are amplitudes of an $ \mathcal{N} = 2 $ topological string on a Calabi-Yau background, see \cite{BCOVReview} and references therein. These amplitudes satisfy a holomorphic anomaly equation which allows for them to be constructed recursively.
		
		In its simplest avatar, the $ 2 $d/$ 4 $d correspondence relates the $ 4 $-point spherical block of a two-dimensional conformal field theory to the instanton partition function of a four-dimensional $ \mathcal{N} = 2 $ supersymmetric gauge theory with gauge group SU$ (2) $ and $ N_f = 4 $ fundamental hypermultiplets \cite{agt}. Another incarnation of the same correspondence establishes a relation between the $ 1 $-point torus block and instanton partition function of an $ \mathcal{N} = 2 $ supersymmetric gauge theory with gauge group SU$ (2) $ and a massive adjoint hypermultiplet \cite{Fateev:2009aw}.\footnote{This is a mass-deformed $ \mathcal{N} = 4 $ supersymmetric gauge theory, and is also referred to as the $ \mathcal{N} = 2^{\star} $ theory.} In particular, the holomorphic anomaly equation relevant to these superconformal gauge theories, studied for example in \cite{Huang:2012kn,1302,1307}, fixes the anomalous modular transformation properties of the prepotential.
		
		We are now in a position to weave together the two threads running through our introduction --- conformal blocks of two-dimensional CFTs and deformed instanton partition functions of supersymmetric gauge theories --- together. Specifically, we can bring to bear results governing instanton expansions in gauge theories on the conformal blocks we're interested in. We observe that the Virasoro conformal blocks are related, via the $ 2 $d/$ 4 $d correspondence, to the deformed prepotential of the appropriate supersymmetric gauge theory. Since the latter is constrained (or recursively determined) by the modular anomaly equation, it must be that the blocks themselves exhibit such a recursive structure. We show that this is indeed the case and the quasimodular structure of the blocks is non-perturbatively captured by the KPZ differential equation or the diffusion equation.

\subsubsection*{Outline}
		In \S \ref{section:2}, after a brief introduction to Zamolodchikov recursion, we determine the subleading contribution in large-$ h_p $ to the block and find that it can indeed be written as a polynomial in the Eisenstein series. We perform various checks of our results, making contact with known exact results. We then comment on the regime of validity of our results, and also point out that there are more profitable ways to rewrite the conformal block that make quasimodular structures more apparent. Finally, we motivate the existence of a modular anomaly equation by studying the fusion kernel and the crossing equation.
		
		In \S \ref{section:3} we elaborate on the $ 2 $d/$ 4 $d correspondence in detail, and establish that the blocks, when expanded in an appropriate combination of the intermediate exchange dimension and central charge, do exhibit this recursive structure and are constrained to satisfy a modular anomaly equation. {Our results in this section are largely inspired by the developments in the study of four-dimensional gauge theories in \cite{1302,1307}.} Finally, an algorithm that recursively builds up the block is also presented.
		
		In \S \ref{section:4} we provide an application of our results. We note that in the bootstrap of \cite{Km}, it is desirable to have closed-form expressions of torus blocks in the internal exchange dimension. Using the subleading results for the blocks, we show that one can systematically find corrections to the asymptotic formula for averaged OPE coefficients.
		
		In \S \ref{section:5} we summarise and discuss possible future directions.

\section{Virasoro blocks in the heavy exchange regime} \label{section:2}

In this section we present the results for the torus $ 1 $-point block and the $ 4 $-point block on the sphere in the regime of large intermediate exchanges ($h_p\to\infty$). We shall consider $c>1$ CFTs with Virasoro symmetry. In order to determine the Virasoro blocks, we shall use the recursion relations discovered by Zamolodchikov \cite{ zamu1, zamu2} for $ 4 $-point block and its adaptation to torus 1-point block \cite{Hadasz:2009db}. The recursion relations are based on the observation that when the block is analytically continued as a function of the central charge $c$ or the intermediate dimension $h_p$, it has poles coming from the singularity structure of degenerate representations. Using the former pole structures give rise to the $ c $-recursion while using the latter give rise to the $ h $-recursion.

In the $ h $-recursion, which is utilised in this section, the poles are located at $h_p = h_{p,mn}$, the dimension of the degenerate representation at level $mn$, corresponding to a null state. These null states make the Verma module, of which it is a part, reducible. This happens since the null state is also a primary, and therefore generates its own Verma submodule. Further, the descendants of null states are also null. This in turn implies that the residue at the pole in the $ h $-recursion is proportional to the block itself. However, the intermediate dimension has now changed to the value at the singularity shifted by the descendant level, i.e. to $h_{p,mn} + mn$. This lends a calculable recursive structure to the blocks, allowing for its perturbative determination as an expansion in the elliptic nome, $q=e^{\pi i \tau}$, associated to the pillow coordinates for the $ 4 $-point block, or the elliptic nome, $q = e^{2\pi i \tau}$, associated with the torus $ 1 $-point block.

\subsection{Torus 1-point block}
We consider the $ 1 $-point correlation functions of  primaries on the torus. This has a decomposition in terms of the torus-$ 1 $ point blocks and OPE coefficients as
\begin{align}\label{tor1}
\vev{O_{h,\bh}}_{\tau,\btau} =\tr\left[O_{h,\bh} \, q^{L_0-c/24}\,\bq^{\bL-c/24}\right] = \sum_{p}C_{pOp}\, \cF_{h_p}(q,h,c)\,\Bar\cF_{\bar h_p}(\bq,\bh,c)~. 
\end{align}
Due to translation invariance, the $ 1 $-point function does not have any dependence on the position coordinate and depends solely on the modular parameter.\footnote{Functions of the modular parameter will be indicated directly by the argument $ \tau $ or indirectly via the argument $ q = e^{2\pi i \tau} $ for the torus $ 1 $-point block and $ q = e^{\pi i \tau} $ for the sphere $ 4 $-point block.}

The torus block contains contributions from the Verma module of each intermediate primary $O_{h_p,\bh_p}$. Its $q$-series can be constructed by calculating the expectation values of the primary $O_{h,\bh}$ in descendants of $O_{h_p,\bh_p}$. We denote a general descendant state as $\ket{\nu_{h_p,N,\{n_i\};\bh_p,\bar N,\{\bar n_i\}}}$, where $N$ and $\bar N$ correspond to a specific integer partition. That is,
\begin{align}
\ket{\nu_{h_p,N,\{n_i\};\bh_p,\bar N,\{\bar n_i\}}} = \prod_{i=1}^{\infty} (L_{-i})^{n_i}(\bL_{-i})^{\bar n_i}\ket{\nu_{h_p;\bh_p}} \ ,
\end{align}
and the expectation value of the external primary $O_{h,h}$ in this state is 
\begin{align}\label{rhorho}
\bra{\nu_{h_p,M,\{m_i\};\bh_p,\bar M,\{\bar m_i\}}}& O_{h,\bh}\ket{\nu_{h_p,N,\{n_i\};\bh_p,\bar N,\{\bar n_i\}}} \\
 &=  C_{pOp} \, \rho\left(h_p,N,\{n_i\};h;h_p,M,\{m_i\}\right)\rho\left(\bar h_p,\bar N,\{\bar n_i\};\bar h;\bar h_p,\bar M,\{\bar m_i\}\right) \ . \nn 
\end{align}
This expectation value needs to be normalized by the inner product of the descendant states, i.e.~the Gram matrix
\begin{align}
\left[B_{h_p,\bar h_p}\right]_{M,\{m_i\};N,\{n_i\}}= \left\langle{\nu_{h_p,M,\{m_i\};\bh_p,\bar M,\{\bar m_i\}}}\big\vert \nu_{h_p,N,\{n_i\};\bh_p,\bar N,\{\bar n_i\}}\right\rangle
\end{align}
Therefore, the $q$-expansion of the conformal block is given by
\begin{align}
\cF_{h_p}(q,h,c) = q^{h_p-c/24}\sum_{n=0}^{\infty} q^n\sum_{\{n_i\}} F_{h_p}^{N,\{n_i\}} \quad \text{with} \quad F_{h_p}^{N,\{n_i\}} = \frac{\rho\left(h_p,N,\{n_i\};h;h_p,N,\{n_i\}\right)}{\left[B_{h_p,\bar h_p}\right]_{N,\{n_i\};N,\{n_i\}}}~. 
\end{align}
Although the $q$-series can be obtained systematically by computing the inner products above, a computationally faster means to achieve the same is to use the recursive representation of the block. In this representation, the block is written as 
\begin{align}
\cF_{h_p}(q,h,c) = \frac{q^{h_p-\frac{c-1}{24}}}{\eta(q)} \cH(h_p,h,c,q)~. 
\end{align}
Here,  $\eta(q)$ is the Dedekind $ \eta $-function and the combination ${q^{h_p-\frac{c-1}{24}}}/{\eta(q)}$ is the character of a primary with conformal dimension $h_p$. The factor $\cH(h_p,h,c,q)$ is to be determined recursively. We avoid repeating the details of the recursion process here and refer the reader to the original work  \cite{Hadasz:2009db}. 

We use the recursion for $\cH(h_p,h,c,q)$ and reorganize the $q$-series into a large-$h_p$ expansion. It can be checked to sufficiently high orders in the $q$-series that each order in the large-$h_p$ expansion can be resummed into polynomials in the Eisenstein series.\footnote{Practically, the specific expressions at each order in $1/h_p$ can be derived by making an ansatz involving a linear combination of the Eisenstein series and its products and then comparing this with the $q$-series from the recursive representation of the block.} The first few orders are shown below
\begin{align}\label{ht2}
{\cal H}_{h_p}(q)&= \ 1  +  \frac{ h(h-1)}{2h_p} \left[\frac{1- E_2(q)}{24} \right]+ \frac{h(h-1)}{8 h_p^2} \bigg\{  ({(h-2)(h-5)-c})\bigg[{1-E_2(q)\over  24}\bigg]^2 \nn \\
&\qquad\qquad+(2h+c-4) \bigg[ \frac{3}{320} - \frac{E_2(q)}{96} - \frac{E_4(q) }{1440} + \frac{E_2(q)^2}{576} \bigg] \bigg\} +  {\mathcal{O}}\left(1/h_p^3\right)  .
\end{align}
The first order correction was derived analytically in \cite[eq.~(80)]{Km} (although it wasn't written in terms of Eisenstein series). The result for higher orders in the large-$h_p$ expansion is provided in Appendix \ref{appB}; it can be seen that each order in the large-$h_p$ expansion can be written in as a linear combination of products of Eisenstein series.  The above expression can also be seen to be correct to very high orders in the $q$-expansion by using numerical values for the external dimension $h$ and the central charge $c$, and then comparing it with the results from the recursive algorithm. The emergence of these quasimodular forms is striking. However, at the same time it is somewhat unusual that the linear combinations involve (quasi)modular forms of different weights. We shall return to this point shortly. 

There are a few immediate checks of the result \eqref{ht2} courtesy of the special points in parameter space where the block is known exactly. For the external operator being the identity, $h=0$, the torus block is just the character and we see from \eqref{ht2} that the $1/h_p$ expansion terminates. Similarly for $h=1$, it can be seen from conformal invariance that the $\rho$ factors in \eqref{rhorho} are 1, i.e.~the expectation value in any descendant state is given by the OPE coefficient itself.\footnote{This can be easily seen by considering the external operator to be a conserved current $J$ which has $h=1$. For the diagonal elements of \eqref{rhorho} the only non-vanishing contribution is from the zero-mode, $J_0$, and this commutes with all $L_n$ modes.} Therefore, the large-$h_p$ expansion terminates in this case as well and the block is given exactly by the character.  Additionally, the expression at $\cO(1/h_p^2)$ passes three tests. It vanishes for $c=0,h=2$ and $c=-2,h=3$ --- these cases have been studied in \cite{Beccaria:2016nnb} and it is known that the block terminates at order $1/h_p$. Finally, our result \eqref{ht2} correctly reproduces the $1/h_p^2$ term for yet another exactly known block at $c=1,h=4$ \cite{Nemkov}. 

\subsection{Sphere 4-point block}
We now consider the Virasoro blocks for the $ 4 $-point function on the sphere/plane. For simplicity, we shall restrict to the special case of the correlator of identical primaries.  Just like the torus $ 1 $-point correlator, the $ 4 $-point correlator can be expanded in a sum over Virasoro blocks as 
\begin{align}
\vev{O(0)O(z,\bz)O(1)O(\infty)}= \sum_{h_p,\bar h_p} C_{OO p}^2 \, \cV_{h_p}(z)\, \bar\cV_{h_p}(\bz)
\end{align}
As alluded to earlier, the Virasoro blocks are not known in closed form but can be determined recursively. The recursion is usually performed in the coordinate $ q $ adapted to the ``pillow'' geometry, given by the orbifold $\mathbb{T}^2/\mathbb{Z}_2$ \cite{Maldacena:2015iua}. The locations of these operators lie at the fixed points of this orbifold. The coordinate on the plane, $z$, is related the $q$ coordinate as
\begin{align}
q=e^{\pi i \tau} \quad \text{with} \quad \tau = i \frac{K(1-z)}{K(z)}~,
\end{align}
while the inverse relation is given by $z={\vartheta_2(q)^4}/{\vartheta_3(q)^4}$. It can then be seen that crossing symmetry, which sends $z\leftrightarrow 1-z$, maps to an S-modular transformation, which sends $\tau \leftrightarrow -1/\tau$. Much like the torus correlator \eqref{tor1}, the $ 4 $-point function in the pillow frame transforms covariantly as a modular form under S-modular transformations. In the recursive representation, the Virasoro block has the form 
\begin{align}\label{Vir-block}
\cV_{h_p}(z) = (16q)^{h_p-\frac{c-1}{24}} [z(1-z)]^{\frac{c-1}{24}-h} \vartheta_3 (q)^{\frac{c-1}{2}-8h} \cH(c,h,h_p,q)~. 
\end{align}
The crucial factor here is $\cH(c,h,h_p,q)$, which is evaluated recursively. For the details of the recursion and some recent developments the reader is invited to consult \cite{zamu1, Perlmutter:2015iya,Chen:2017yze,Kusuki:2018nms}.

Using the recursion relations, it can be seen that the $q$-expansion can once again be organized into a linear combination of the Eisenstein series and its various products.
\begin{align} \label{H-next-2}
&\cH(c,h,h_p,q) \nn \\
=& ~1 - \frac{1}{16 h_p }\left( (c+1)-32 h \right)\left( (c+5) - 32 h  \right)\left[ \frac{E_2(q)-1}{24} \right]\nn \\
&+\frac{1}{h_p^2}\bigg\{-\tfrac{ \left(4 h -1\right)  {\left(c-32 h +1\right) \left(c-32 h +5\right)}}{1152}E_2(q)\nn +\tfrac{ {\left(c-32 h +1\right) \left(c-32 h +5\right)} \left(c-32 h +9\right) \left(c-32 h +13\right) }{512} \left[\frac{E_2(q)-1}{24}\right]^2\nn \\
&\qquad\quad+\tfrac{\left(32 h  \left(-32 h  \left(c+32 h -41\right)+c (5 c-58)-143\right)+c (c (17-3 c)+111)+115\right)}{92160}E_4(q)\nn \\
&\qquad\quad+\tfrac{32 h  \left(32 h  \left(-19 c+352 h -181\right)+c (5 c+278)+673\right)+c (c (3 c-97)-591)-515}{92160} \bigg\} + \cO\left(1/h_p^3\right)~. 
\end{align}
The appearance of $E_2(q)$ at leading order was found most recently in \cite{Cardona:2020cfy}. However, it is observed that similar structures appear at higher orders in the large-$h_p$ expansion as well.  As before, the above expression in terms of the Eisenstein series can be checked numerically to high orders in the $q$-expansion by using numerical values $h$ and $c$. The expression for the next order in large-$h_p$ expansion can be found in Appendix \ref{appB}. 

Exact results for blocks associated to the $ 4 $-point correlator are fewer in number than those known exactly for the torus $ 1 $-point case. The only exact solutions known (for the case of equal external dimensions) are the ones for $c=1,h=1/16$ and $c=25,h=15/16$. In both these cases, one has $\cH(c,h,h_p,q)=1$, i.e.~the expansion terminates at the zeroth order. This can be seen to happen for the first two orders presented in \eqref{H-next-2} and we have also verified the same at higher orders.

We observe that for external dimensions, $h = h_* \in \left\lbrace \tfrac{c+5}{32}, \tfrac{c+1}{32} \right\rbrace$ the first order correction in \eqref{H-next-2} vanishes. Also the first two terms of the $1/h_p^2$ piece in \eqref{H-next-2} vanishes. In fact for $h = h_*$, the $1/h_p^{2}$ coefficient admits a fairly simple form:
\begin{align}\label{H-next}
\cH(c,h_* ,h_p,q) 
&= 1 +  \frac{c_* - c}{32 h_p^2} \left[\frac{ E_4(q) -1 }{240}\right] + \cO\left(1/h_p^3\right) \nn 
\end{align}
where, for $h_* = \tfrac{c+5}{32}$, $c_*=25$ while for $h_* = \tfrac{c+1}{32}$, $c_*=1$.

\subsection{Further comments}\label{S:Comments}
\begin{figure}[t]
	\centering
	\includegraphics[width=\linewidth]{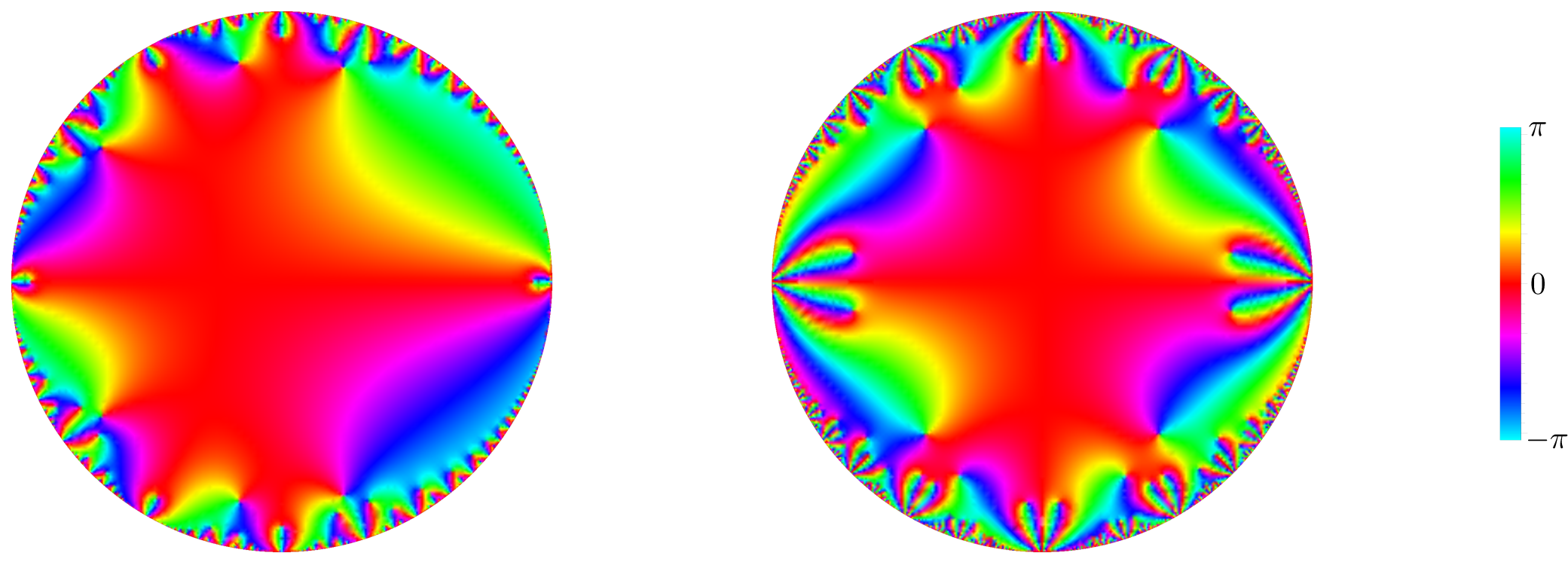}
	\caption{\textbf{Plotting $\arg[\cH(q)]$ ---} [Left] for the 1-point torus block till $\cO(1/h_p^6)$ plotted on the unit-disk with $c=1,h=4,h_p=6$;  and [Right] for the  4-point block on the sphere till $\cO(1/h_p^4)$ with the parameters $c=3,h=0.8,h_p=8$.}
	\label{fig:blockdiscs}
\end{figure}

As alluded to earlier,  knowing the blocks as polynomials of the Eisenstein series in the $1/h_p$ expansion specifies them completely on the unit-disk in the $q$-plane. This is demonstrated in Figure \ref{fig:blockdiscs}, where we plot the recursion factor $\cH$ for the blocks considered. For the   torus block we have chosen parameters, $c=1$ and $h=4$, to compare with the block known exactly \cite[eq.\,(36)]{Nemkov}. The plot of the exact result shows almost no visible differences with the $1/h_p$-expanded recursive factor to the sixth order. 

In the following we further discuss the modular aspects of the expressions obtained for the blocks and motivate developments for the next sections.

\subsubsection*{Regime of validity}
It is important to understand what the regime of validity of our results is.\footnote{We thank Sridip Pal for raising this question.} In order to determine this, we note that \emph{all} terms in the $q$-series are important in the (high-temperature) limit $q\to 1$ or $\im \,\tau \to 0^+$. The other parameters of the system --- the central charge $c$, and the external dimension $h$ --- are being held fixed.  We can use the S-duality relations \eqref{mod-prop-Ek} for the Eisenstein series
to find the behaviour of the leading correction in \eqref{ht2} or \eqref{H-next-2} in the $q\to 1$ regime:
\begin{equation}\label{regime}
 \frac{E_2(\tau)-1}{h_p} =  \frac{\tau^{-2}E_2(-1/\tau)- \tfrac{6}{i\pi \tau }-1}{h_p} \ \stackrel{\tau\to i0^+}{\approx}  \ \frac{1}{h_p\tau^2}~. 
\end{equation}
The same conclusion can be reached by using the lattice sum \eqref{lattice-Ek}. 
Hence the first order correction is small provided \begin{align} \label{condition} 
h_p|\tau|^2 &\gg 1, 
\end{align} 
or $h_p \gg 1/|\tau|^2$. This is the necessary condition for which the large-$ h_p $ expansion can be trusted. 

Let us define the quantity $\delta\equiv(h_p\tau_2^2)^{-1}$, in terms of which \eqref{condition} translates to $ \delta \ll 1 $. It can also be seen that the higher order terms scale as $\delta^n$. A similar conclusion was reached from a slightly different perspective in \cite{Kusuki:2018wcv}.  We can view this restriction via the unit disk in the complex plane of $q=e^{2\pi i \tau}$. The limit $\im \,\tau = \tau_2 \to 0$ corresponds to the edge of the disk.  The condition \eqref{condition} implies that the analysis can be trusted anywhere within a slightly smaller radius of  $ \exp\left({-{2\pi}\sqrt{\delta/h_p}}\right)\approx 1 -{2\pi}\sqrt{\delta/h_p}$\,.

\subsubsection*{A double scaling limit}
On eyeballing \eqref{ht2} and \eqref{H-next-2}, it appears plausible that a judicious double scaling limit will further simplify the block.\footnote{We are grateful to Alexander Zhiboedov for encouraging us to explore this possibility.}  Consider the limit of the block in which $
h, h_p \rightarrow \infty$ with the ratio $h^2/h_p = \kappa$ held fixed. Indeed, as we will demonstrate, the block can be determined in a closed form in this regime.

For the torus $ 1 $-point block, this double scaling limit yields
\begin{equation}\label{key}
\begin{aligned}
\cH_{\mathbb{T}^2} &= 1+\frac{\kappa}{48} \left[1-E_2(q)\right] + \frac{\kappa^2}{4608}  \left[1-E_2(q)\right]^2 + \cdots \ ,
\end{aligned}
\end{equation}
while for the sphere $ 4 $-point block, we have
\begin{equation}\label{key}
\begin{aligned}
\cH_{\mathbb{S}^2} &= 1+\frac{8\kappa}{3} \left[1-E_2(q)\right] + \frac{32\kappa^2}{9}  \left[1-E_2(q)\right]^2 + \cdots \ .
\end{aligned}
\end{equation}
In both cases, since we have held the central charge $ c $ fixed, it has dropped out of the final expression. While we have only presented the first two orders in the $ \kappa $-expansion, one can check that the above form persists at higher orders and the above expansions can be resummed into :
\begin{equation}\label{key}
\begin{aligned}
\cH_{\mathbb{T}^2} \approx \exp \left\lbrace \frac{h^2}{48h_p} \left[1-E_2(q)\right] \right\rbrace \ , \qquad
\cH_{\mathbb{S}^2} \approx \exp \left\lbrace \frac{8h^2}{3h_p} \left[1-E_2(q)\right] \right\rbrace \ .
\end{aligned}
\end{equation}
It is interesting that in this double scaling limit, the recursion factor of the block exponentiates.

\subsubsection*{The Liouville parametrization}

We have observed in \eqref{ht2} and \eqref{H-next-2} that the `recursive factor' $\cH$  of the Virasoro block for both the sphere and the torus organizes itself into polynomials of the Eisenstein series in the $1/h_p$ expansion. However, each order in this perturbative expansion does not have definite modular properties as we have a mix of terms of varying modular weights. It turns out that using the Liouville parametrization for the intermediate conformal dimension furnishes a better reorganization of the expansion.  In this parametrization, the central charge and the intermediate conformal dimension are
\begin{align}
\label{liouvilleparametrisation}
c = 1 + 6 Q^2 \quad \text{and} \quad h_p = \frac{Q^2}{4} - \alpha^2 \ . 
\end{align}
 Sometimes, $\alpha$ is also referred to as momentum since it appears as the momentum of the vertex operator. One motivation to consider an expansion of the blocks in large-$\alpha$ is that matrix elements of light primaries in descendant states of other heavy primaries have a natural expansion in this large parameter \cite{Besken:2019bsu}. The Virasoro block repackages this information about these matrix elements.

It turns out that the logarithm of $\cH$, and not $\cH$ itself, has the cleanest modular features. Consequently, we will reorganize the large-$h_p$ expansion of the blocks into a large-$\alpha$ expansion of the logarithm of the recursive part of the blocks.
\begin{align} \label{frei-energie}
&{\cal H}(q) = 1 + \sum_{n=1}^\infty \frac{ {\cal A}_n (q) }{ h_p^n } = e^{-F(\alpha,q)} = \exp \left( \sum_{j=1}^\infty \frac{A_j (q) }{\alpha^{2j}} \right) \ ,\\
&\text{with, }A_j(q) = {\cal K}_j + \frac{\tilde{h}_j (q)}{ 2^{j+1}\,j } \ .  \nn 
\end{align}
In the above equations, if $ \cH $ is likened to a partition function, then $F(\alpha,q)$ is analogous to the free energy, so we will occasionally refer to it as such.\footnote{This is like studying the effective action, which has contributions only from connected diagrams.} The separation of $A_j(q)$ into quasimodular $ (\tilde{h}_j) $ and constant $ (\mathcal{K}_j) $ pieces   will be important in what follows. It is helpful to note at this point that the abstruse normalization factors in this subsection are conveniently chosen to relate to quantities appearing in the forthcoming section. 

The expression for the free energy for the torus $ 1 $-point block is, from \eqref{ht2}
\begin{align}\label{F-torus}
F_{\mathbb{T}^2}(\alpha,q)  = \frac{h(1-h)}{48} \bigg[ &
-\frac{1}{\alpha^2}+\frac{ 12 h + c  -19}{240 \alpha^4} + \cdots  \\
&+ \frac{E_2(q) }{\alpha^2} + \frac{1}{\alpha^4}\bigg[ \frac{ ( 4-c-2h) }{240  }E_4(q)+\frac{ 5 (3-2h) }{240 }E_2(q)^2\bigg] + \cdots \bigg]\nn . 
\end{align}
In the two lines above, we have separated out the large-$\alpha$ expansion into pieces that are independent of $ q $ and pieces that are quasimodular.\footnote{The part independent of $\tau$ is in many ways fictitious, as the terms get canceled by the $q\to 0$ limit of the Eisenstein series.} It can be seen from the second line that coefficients of $\alpha^{-2n}$ are quasimodular forms of weight $2n$. The same feature can also be seen for the $ 4 $-point block on the sphere:
\begin{align}\label{F-sphere}
F_{\mathbb{S}^2}(\alpha,q) =\,& \frac{1}{\alpha^2}\tfrac{(c+1-32h)(5+c-32h)}{384 } +\frac{1}{\alpha^4} \tfrac{ (145 +  173 c + 41 c^2 + c^3 - 6048 h - 32 c h ( 5c + 94) + 1024 h^2 ( 7c + 53 )  - 98304 h^3 )}{46080 } + \cdots\nn \\
&- \frac{1}{\alpha^2}\tfrac{(c+1-32h)(5+c-32h)}{384 }E_2(q) 
+\frac{1}{\alpha^4}\bigg[\tfrac{  (32h - c -1)(5+c-32h) ( 7 + c - 32 h) }{18432 }E_2(q)^2
 \\ 
&\hspace{2cm}+  \tfrac{( 4576 h - 115 -111 c -17 c^2 + 3 c^3 + 32 c h( 58-5c) + 1024h^2(c-41) + 32768h^3 )}{92160} E_4(q)\bigg] + \cdots .\nn 
\end{align}
These observations suggest that the free energy has modular transformation properties that are more amenable to analysis.

\subsubsection*{Hints of recursion from the fusion kernel}
We can see from the first few orders in the large-$\alpha$ expansion that the coefficients $\tilde{h}_j$ in \eqref{frei-energie} have a modular anomaly courtesy of their dependence on the quasimodular form $E_2(q)$. This implies that the modular S-transformation acts as 
 \begin{align}
 \label{S-trans}
 {\rm S} \left[ \tilde{h}_j \right] &= \tau^{2j} \left(\tilde{h}_j + \sum_{k=1}^\infty \frac{1}{k!} \frac{1}{ ( 2\pi i \tau )^k } (12 \partial_{E_2})^k \tilde{h}_{j} \right) . 
 \end{align}
 The second term above is a consequence of quasimodularity. 
 Let  us consider the case of the $ 1 $-point torus block --- similar arguments apply for the $ 4 $-point block on the sphere as well. 
 The coefficients $\tilde{h}_j$ are further constrained by the crossing equation that the block needs to satisfy. For the torus block one has, 
\begin{align}\label{crossing}
\frac{e^{-2 \pi i \tau \alpha^2}}{\eta(\tau)} {\cal{H}}_\alpha (q) &= \int_{-\infty}^\infty d\alpha'\,\,M_{\alpha,\alpha'} \frac{e^{\tfrac{2\pi i \alpha'^2}{\tau} }}{\eta(-1/\tau)} {\cal{H}}_{\alpha'} ( \tilde{q} )~. 
\end{align} 
In the above, $M_{\alpha, \alpha'}$ is the fusion kernel for which an explicit form exists \cite{Nemkov2}.\footnote{For the 4-point block the explicit forms are derived in \cite{Ponsot:1999uf, Ponsot:2000mt}.} Note that the modular S-transformation effects the exchange $\alpha \leftrightarrow \alpha'$. It has been conjectured in \cite{Galakhov:2012gw, Nemkov1307} that perturbatively in large-$\alpha$, the S-modular transformation is \emph{exactly} the Fourier transform.  In \cite{Galakhov:2012gw} this was argued for by explicit calculation of the deformed block using matrix model technology but for specific parameters. In \cite{Nemkov1307} the modular kernel computation was generalized for arbitrary external operators and central charge till order $\CO\left(\alpha^{-6}\right)$ and on the heels of this observation, the aforementioned conjecture was made. Assuming this simplified form of the kernel, one can show that S$^2[\alpha] = -\alpha$. The leading saddle-point in the momentum integral of the crossing equation \eqref{crossing} gives 
\begin{align}
\label{saddle}
{\rm S}[\alpha] = \alpha' \approx - \alpha \tau - \frac{1}{4\pi i } \frac{\partial  F(\alpha,q)}{\partial \alpha }.
\end{align}
where we used the definition in \eqref{frei-energie}. We now act with a modular S-transformation on the above equation and use the explicit transformation \eqref{S-trans}, along with the result S$^2[\alpha] = -\alpha$. We find
\begin{align}\label{whatis}
\sum_{k=1}^\infty\,\, \frac{12 \partial_{E_2} \tilde{h}_k}{ 2^k}  \frac{1}{\alpha^{2k+1}} -  \sum_{k,m=1}^\infty\,\, \frac{(2k+1)\tilde{h}_k \tilde{h}_m }{2^{k+m+1} }  \frac{ 1 }{ \alpha^{2m+2k+3} } &\approx 0 \ .
\end{align}
Finally, comparing powers of $\alpha$ yields the following recursion relation
\begin{align}\label{recrec1}
 \frac{\partial \tilde{h}_\ell }{\partial E_2 } &\approx \frac{\ell}{12} \sum_{i=0}^{\ell - 1} \tilde{h}_i \tilde{h}_{\ell - i -1} ~. 
\end{align}

It is easy to check at low orders in the large-$ \alpha $ expansion that this recursion relation is almost (but not quite) satisfied. Nevertheless, it provides a plausible mechanism for how quasimodular forms are generated at each order in the large-$\alpha$ expansion. Similarly, the fusion kernel for the $ 4 $-point block on the sphere differs by a factor of $ 2 $ in the Fourier exponent from the torus $ 1 $-point block kernel. This is reflected in the analogue of equation \eqref{recrec1} which now reads
\begin{equation}\label{recrecfootnote}
 \frac{\partial \tilde{h}_\ell }{\partial E_2 } \approx \frac{\ell}{6} \sum_{i=0}^{\ell - 1} \tilde{h}_i \tilde{h}_{\ell - i -1} ~. 
\end{equation}
We emphasise that the simplification of the Virasoro crossing kernel to the Fourier kernel \cite{Galakhov:2012gw, Nemkov1307} in the large momentum expansion is an observed fact, and it is  unclear as yet how to directly relate the non-perturbative corrections to the kernel of the blocks \cite{Nemkov2}.


To summarise, in this section we have observed the emergence of quasimodular forms at all orders in the large-$ h_p $ expansion of the Virasoro blocks and have tried to explain this observation from a purely CFT standpoint. This story is, however, incomplete as we need to go beyond the saddle point approximation and a direct approach seems beyond reach. Certain tantalising similarities with structures more commonly associated to four-dimensional supersymmetric gauge theories, however, move us to look to the $ 2 $d/$ 4 $d correspondence for further clues. A discussion of these aspects is the subject of the Section \ref{section:3}. 

Before we discuss the import from gauge theory, however, we take a brief detour and discuss a simple yet striking example --- essentially a ``toy'' version of our analysis of the conformal block --- that contains within it a useful blueprint for what structures to expect.

	\subsubsection*{Intermezzo: Thetas and toy blocks}

We now present a simple example of a mathematical function where quasimodular structures appear in a strikingly similar fashion, with much of the structure exhibited by the conformal block and none of the complications. As with all toy examples, the analogies presented in this section should be taken with a pinch of salt, since the recursion factor $ \cH $ is far more intricate.

Consider the Jacobi theta function,
$
\vartheta_1(\tau,z)
$. 
This object can be written in terms of the Eisenstein series as \cite[eq.\,(20.6.2)]{DLMF}
\begin{align}
\vartheta_1(\tau,z) = z \,\eta(\tau)^3 \exp\left[\sum_{k=1}^{\infty}z^{2k}\frac{(-1)^kB_{2k}}{2k(2k)!} E_{2k}(\tau)\right]~,
\end{align}
where $ B_{2k} $ are Bernoulli numbers. The form of the exponential bears a close resemblance to \eqref{frei-energie}. That is, (quasi)modular forms of definite weight appear at each order in the $z$-expansion of $\log\vartheta_1(\tau,z)$. By comparing the two expressions, we read off that the analogue of the $z$-variable in the Virasoro conformal block is $1/\alpha$.  Note, however, that in this case $z$ is the elliptic variable and it has a much simpler modular transformation property: $z\mapsto (c\tau+d)^{-1}z$. Under the S-modular transformation, $\vartheta_1$ transforms as 
\begin{align}
\vartheta_1\left(z/\tau, -1/\tau\right) = e^{\frac{\pi iz^2}{\tau}} (-i\tau)^{1/2} \vartheta_1(z,\tau)~. 
\end{align}
This is essentially an analogue of the crossing equation \eqref{crossing} with the fusion kernel $M_{z z'}=\delta(z-z')$. 
The parallels here are quite striking, and one might be tempted to take them more seriously. For example, note that the function $ \vartheta_1(\tau,z) $ satisfies a heat equation:
\begin{equation}\label{key}
\left(\frac{\partial}{\partial \tau} + \frac{i}{4\pi} \frac{\partial^2}{\partial z^2} \right) \vartheta_1 (\tau,z) = 0 \ .
\end{equation}
If we were to take this analogy seriously, one might expect that heat equations would constrain the conformal block as well. The developments in the following section, which draw on intuition from the literature on supersymmetric gauge theories, will allow us to see that this is precisely the case.

\section{Modular features and the 2d/4d correspondence}\label{section:3}

In the previous section, we saw that reconstituting the large-$ h_p $ expansion of the block into a large-$ \alpha $ expansion was profitable from two different points of view. First, we saw that this expansion of the logarithm of the block had coefficients with definite modular weight. Second, we observed that the momentum $ \alpha $ appeared naturally in the fusion kernel, and on taking the large-$ \alpha $ limit, we were able to derive a constraint satisfied by the coefficients $ \tilde{h}_k $. It is also curious that the relation defining the saddle-point of the crossing equation \eqref{saddle} looks strikingly similar to the relation between periods, dual periods, and the prepotential of four-dimensional $ \mathcal{N} = 2 $ supersymmetric gauge theories. 

In this section, we will argue that \eqref{recrec1} and \eqref{recrecfootnote} need to be corrected, and that the form of this correction can be divined by appealing to the $ 2 $d/$ 4 $d correspondence. We will then use this to provide a faster algorithm for computing the conformal block in the heavy exchange regime. 

\subsection{Parameter maps}

As we alluded to in the introduction, the $ 2 $d/$ 4 $d correspondence establishes an equivalence between the instanton partition function of an $ \mathcal{N} = 2 $ supersymmetric gauge theory and the the conformal block. The cases of interest in this paper --- the $ 4 $-point block on the sphere with equal external dimensions, and the $ 1 $-point block on the torus --- each correspond to a $ \Omega $-deformed supersymmetric gauge theory with specific matter content. For convenience, we write the $ \Omega $-deformation parameters in terms of their sum and product as
\begin{equation}\label{key}
s = \epsilon_{1} + \epsilon_{2} \quad \text{and} \quad p = \epsilon_{1}\epsilon_{2} \ .
\end{equation}
As an aside, in the refined topological string \cite{Iqbal:2007ii}, the quantity $ p $ is simply the string coupling constant. This way, according to \cite{agt}, the number $ Q^2 $ that parametrises the central charge as in \eqref{liouvilleparametrisation} in the dual conformal field theory is given by
\begin{equation}\label{key}
Q^2 = \frac{s^2}{p} \ .
\end{equation}
In the following, we will lay out the specific parameter map in either case.

\subsubsection*{1-point block on the torus}

The $ 1 $-point block on the torus, via the $ 2 $d/$ 4 $d correspondence, maps onto an $ \Omega $-deformed SU$ (2) $ gauge theory with an adjoint hypermultiplet of mass $ m $. The map of parameters is:
\begin{equation}\label{key}
h = \frac{Q^{2}}{4}-\frac{m^{2}}{p} \quad \text{and} \quad h_p = \frac{Q^{2}}{4}-\frac{a^{2}}{p} \ .
\end{equation} 
On the left-hand side of the above equations, we have the conformal dimensions of the external $ (h) $ and exchanged $ (h_p) $ operators, while on the right-hand side we have parameters in the gauge theory: the mass $ m $ of the adjoint hypermultiplet and the Coulomb vacuum expectation value $ a $ of the adjoint scalar in the $ \mathcal{N} = 2 $ vector multiplet. 

\subsubsection*{4-point block on the plane}

The $ 4 $-point block on the sphere with equal external dimensions, via the $ 2 $d/$ 4 $d correspondence, maps onto an $ \Omega $-deformed SU$ (2) $ gauge theory with $ N_f = 4 $ fundamental hypermultiplets. The map of parameters is:
\begin{equation}\label{key}
h = \frac{Q^{2}}{4}-\frac{m^{2}}{4p} \quad \text{and} \quad h_p = \frac{Q^{2}}{4}-\frac{a^{2}}{p} \ .
\end{equation} 
On the left-hand side of the above equations, we have the conformal dimensions of the external $ (h) $ and exchanged $ (h_p) $ operators, while on the right-hand side we have parameters in the gauge theory. As before, $ a $ is the Coulomb vacuum expectation value of the scalar in the adjoint vector multiplet. For the four fundamental masses $  m_i $, we start with the more general map relevant for arbitrary external operator dimensions $  h_i $:
\begin{equation}\label{key}
\begin{aligned}
h_1 = \frac{Q^{2}}{4}-\frac{(m_1 - m_2)^{2}}{4p} \quad \quad h_2 = \frac{Q^{2}}{4}-\frac{(m_1 + m_2)^{2}}{4p} \ ,\\
h_3 = \frac{Q^{2}}{4}-\frac{(m_3 - m_4)^{2}}{4p} \quad \quad h_4 = \frac{Q^{2}}{4}-\frac{(m_3 + m_4)^{2}}{4p} \ . \\
\end{aligned}
\end{equation}
In order to recover the case of equal external dimensions, we choose $ m_1 = m_4 = 0 $ and $ m_2 = m_3 = m $.  Now that the parameter maps are explicit, we are in a position to discuss the recursion relation that will be relevant to the conformal blocks.

\subsection{From modular anomalies to the diffusion equation}
First, observe that if we work in units where $ p = 1 $, we recover the Liouville parametrisation we introduced earlier, i.e.~the Liouville momentum $ \alpha $ is identified with $ a $. This is simply a convenient parametrisation, and represents no loss of generality, since the factors of $ p $ can be restored by dimensional considerations.

Second, the constraint \eqref{recrec1} also arises in supersymmetric gauge theories, when attempting to constrain the dependence of the prepotential on the weight-$ 2 $ quasimodular Eisenstein series \cite{Billo:2015pjb,Billo:2015jyt}. In these papers, however, the gauge theories are undeformed. Since the $ 2 $d/$ 4 $d correspondence requires deformations on the gauge theory side, one might hope that an analogue of \eqref{recrec1} exists for the deformed gauge theories as well. Fortunately, this has been done in the deformed gauge theory context in \cite{1302}, so let us review their arguments.

Recall the amplitudes $ F^{(n,g)} $ from the introduction. The modular anomaly equation that constrains their dependence on the quasimodular weight-$ 2 $ Eisenstein series is
\begin{equation}\label{key}
\partial_{E_{2}} F^{(n, g)}=-\frac{1}{24 M} \sum_{i=0}^{n} \sum_{r=0}^{g} \partial_{a} F^{\left(i, r\right)} \partial_{a} F^{\left(n-i, g-r\right)}+\frac{1}{24 M} \partial_{a}^{2} F^{(n, g-1)} \ ,
\end{equation}
where $ M = 1 $ for the $ N_f = 4 $ theory (the spherical block) and $ M = 2 $ for the $ \mathcal{N} = 2^{\star} $ theory (the torus block). This equation is equivalent to the holomorphic anomaly equation \cite{Bershadsky:1993ta,Bershadsky:1993cx}, and in that context these two terms correspond to the two possible degenerations of a genus-$ g $ Riemann surface: into two surfaces of genus $ (r) $ and $ (g-r) $, and into a surface of genus $ (g-1) $ after a cycle is pinched. Schematically, the term $ \partial F \partial F $ on the right-hand side of the above equation is the analogue of the right-hand sides of \eqref{recrec1} and \eqref{recrecfootnote}. Our analysis of the fusion kernel did not yield a linear term analogous to $ \partial^2 F $, so given the correspondence between two dimensional conformal field theories and four-dimensional gauge theories, it is perhaps natural to guess that an analogue of this term is what we are missing. 

By keeping track of powers of $ \alpha $, it is easy to see that the term proportional to $ \partial^2 F $ will correspond to an additional term modifying \eqref{recrec1} and \eqref{recrecfootnote}, in each case proportional to $ \tilde{h}_{\ell-1} $. Purely based on considerations of modular weight, of course, this term is permitted. We find with some simple algebra that the modular anomaly equations relevant to the conformal block take the form
\begin{align}
\text{Torus 1-Point Block} &\colon \quad \frac{\partial \tilde{h}_\ell }{\partial E_2 } = \frac{\ell}{12} \sum_{i=0}^{\ell - 1} \tilde{h}_i \tilde{h}_{\ell - i -1} + \frac{\ell ( 2\ell -1 ) }{12 } \tilde{h}_{\ell-1}, \label{recrec} \\
\text{Sphere 4-Point Block} &\colon \quad \frac{\partial \tilde{h}_\ell }{\partial E_2 } = \frac{\ell}{6} \sum_{i=0}^{\ell - 1} \tilde{h}_i \tilde{h}_{\ell - i -1} + \frac{\ell ( 2\ell -1 ) }{6 } \tilde{h}_{\ell-1}. \label{recrec2}
\end{align}
It will be convenient to package both these recursions into one, as
\begin{equation}\label{eq:MasterRecursion}
\frac{\partial \tilde{h}_\ell }{\partial E_2 } = \frac{\ell}{6M} \sum_{i=0}^{\ell - 1} \tilde{h}_i \tilde{h}_{\ell - i -1} + \frac{\ell ( 2\ell -1 ) }{6M} \tilde{h}_{\ell-1}
\end{equation}
with $ M = 1 $ for the spherical block and $ M = 2 $ for the torus block. For low orders in the large-$ \alpha $ expansion we computed in Section \ref{section:2}, it can be checked explicitly that the above recursion is satisfied. The seed for the recursion in both cases is
\begin{equation} 
\tilde{h}_0 = \begin{dcases}
 \frac{c+1-32h}{4} \quad &\text{ for sphere block}, \\
 -h \quad &\text{ for torus block}. 
\end{dcases}
\end{equation}
We observe, however, that $ \tilde{h}_0 $ is \emph{not} a part of the block, see \eqref{frei-energie}. At this point, we also remark that the explicit results of the previous section for the block in the $1/\alpha$ expansion are verified to be perfectly  consistent with the recursion  \eqref{eq:MasterRecursion}. 

It is desirable to bind the recursion relations for $\th_\ell$ non-perturbatively into a single partial differential equation. This development is inspired by a study of $ \Omega $-deformed gauge theories due to \cite{1307}.  Consider the function
\begin{equation}\label{eq:F0Definition}
\cF(q) = -\tilde{h}_0 \log 2\alpha + \sum_{j=1}^\infty \frac{\tilde{h}_j (q)}{ 2^{j+1}\,j \,\alpha^{2j}}\ ,
\end{equation}
where $\tilde{h}_j  $ are recursive pieces for the sphere or the torus block the case may be -- cf.~equation \eqref{frei-energie}. Indeed, using the $ 2 $d/$ 4 $d correspondence it is possible to locate the origin of the logarithmic term in \eqref{eq:F0Definition} in the $ 1 $-loop contribution to the prepotential.  It is then straightforward to show that the differential equation
\begin{equation}\label{KPZ}
\partial_{E_{2}} \cF = \frac{1}{24M} \left[ \left(\partial_\alpha \cF \right)^2 + \partial_{\alpha}^2 \cF \right] \ ,
\end{equation}
reproduces the recursions \eqref{eq:MasterRecursion}. This is the one-dimensional noiseless Kardar-Parisi-Zhang equation, with the time proportional to $ E_2 $ and the role of space is played by $ \alpha $. We emphasise that the object $\cF$ is closely related to the Virasoro block, via \eqref{frei-energie}, and the above differential equation provides a non-trivial constraint which suffices to completely fix the purely $E_2 $ dependent part of the block. To make this explicit, we note that the recursive factor of the block can be written as follows
\begin{align} \label{HF-relation}
&{\cal H}(q) = \exp \left( \cF(q) + \th_0 \log 2\alpha +\sum_{j=1}^\infty    {\cal K}_j { \alpha^{-2j}} \right) = \exp \left( \cF(q) - \cF(0) \right) .
\end{align}
using \eqref{frei-energie} and \eqref{eq:F0Definition}. 

On taking a further derivative with respect to $ \alpha $, one finds from \eqref{KPZ} that $ \partial_\alpha\cF $ satisfies the viscous Burgers equation, which by using the Hopf-Cole transformation can be linearised into the heat/diffusion equation. That is
\begin{equation} \label{heat} 
\partial_{E_{2}}  \mathcal{Z} - \frac{1}{24M} \partial_{\alpha}^2  \mathcal{Z} = 0~, \qquad  \cF(q) = \log \mathcal{Z}(q) \ . 
\end{equation}
The quantity $\mathcal{Z}(q)$ is related then to the recursive part of the Virasoro block as $\cH(q)=\mathcal{Z}(q)/\mathcal{Z}(0)$. 
It is conceivable that heat kernel methods for solving \eqref{heat} might  be pressed into the service of investigations into the Virasoro block in  future. At a more fundamental level, it would be valuable to derive \eqref{KPZ} or \eqref{heat} by using an entirely 2d CFT based approach. For now, we turn to a discussion of the modular anomaly equations and how they can be used to recursively construct the block.

\subsection{An algorithm for faster computations of Virasoro blocks }\label{S:Algo+}

In the previous section, after drawing inspiration from the $ 2 $d/$ 4 $d correspondence, we determined that the Virasoro blocks under consideration, when expanded in large-$ \alpha $, have coefficients $ \tilde{h}_{\ell} $ that can be found recursively via  modular anomaly equations: for the torus $ 1 $-point block, we have \eqref{recrec}, and for the sphere $ 4 $-point block, we have \eqref{recrec2}. In this section, we present an algorithm that systematically computes the large-$ \alpha $ (and consequently, a large-$ h_p $) expansion of the block using these anomaly equations.

It is more convenient to work with the logarithm of the block ${\cal{H}}$, and only later exponentiate it to required order in $1/h_p$. In order to keep the notation light, we will suppress the dimensions of the external and exchanged operators, so $ \cH \equiv \cH_{h_p}^{h} $. 

We will start by expressing $F = \log {\cal{H}}$ as a series in large $\alpha^2 = \left(\tfrac{c-1}{24} - h_p\right)$. To do this, begin by defining the partial sums
\begin{equation} 
\label{eq:FmLargeAlpha}
F^{(m)} = \sum^m_{ j \geq 1} \frac{\cL^{(j)}}{\alpha^{2j}} \ . 
\end{equation} 
If all the coefficients $ \cL^{j} $ are determined, we can recover the block by considering the series associated to the partial sums
\begin{equation}\label{key}
\log {\cal{H}} = \lim_{n \rightarrow \infty}F^{(n)} \ .
\end{equation}
Our goal is to construct the large-$ \alpha $ expansion recursively, which means given the partial sum $ F^{(k-1)} $, we want to determine the partial sum $ F^{(k)} $ using the recursion relations.

The algorithm takes as input the partial sum $F^{(k-1)}$, which includes the set $\{ \tilde{h}_j \}_{j=0}^{k-1}$, which will feature in the recursion \eqref{eq:MasterRecursion}. A finite number of terms from the $ q $-series produced by Zamolodchikov's $h$-recursion will also form part of the input in order to fix the purely modular pieces --- this is a direct analogue of the use of Nekrasov's equivariant localisation in the gauge theory deployment of the modular anomaly equation \cite{1302,1307}. The $h$-recursion allows one to get the recursive block (denoted $ \cH_{\text{Z}} $) till order $q^n$, at computational cost growing as $\CO \left( n^3 (\log n)^2 \right)$. We express the logarithm of the same as a large-$ \alpha $ series and it takes the following form,
\begin{align}
\label{data-ZR}
\log {\cal{H}}_{\text{Z}} &=  \sum_{n\geq 1} \frac{1}{\alpha^{2n}} \left( \sum_{m \geq 1} a_m^{(n)} q^m \right).
\end{align}
As we shall see, at a given order $2k$ in the large-$ \alpha $ expansion we shall need the $q$-expansion up to order $d_k$, the dimension of the space of weight-$ (2k) $ modular forms. For low orders in the large-$ h_p $ expansion this is just a handful of coefficients.

Start with the set $\{ \tilde{h}_j \}_{j=0}^{k-1}$ to evaluate the r.h.s.~of \eqref{eq:MasterRecursion}, and integrate this with respect to $E_2$ to find the quasimodular part $ \mathcal{Q}^{(k)} $ of $\tilde{h}_k$: 
\begin{align}\label{e2part}
\mathcal{Q}^{(k)} &= \frac{1}{2^{k+1} k} \int \, \text{d}E_2\,\,\left( \frac{k}{6M} \sum_{i=0}^{k - 1} \tilde{h}_i \tilde{h}_{k - i -1} + \frac{k ( 2k -1 ) }{6M} \tilde{h}_{k-1} \right).
\end{align}  
The above terms fix all $E_2$ dependence at this order in the large-$ \alpha $ expansion, which means it contains all possible pieces of the form $E_2^{m_1} E_4^{m_2} E_6^{m_3}$ such that $2m_1 + 4m_2 + 6m_3 = 2k$, with $m_1 \geq 1$. 

The dimension of modular forms at this order is given by integer partitions of $ k $ using only the integers $\lbrace 2, 3 \rbrace $ which is $d_k = k+1  - \ceil{k/2} - \ceil{k/3}$. Upto $d_k$ undetermined coefficients, $d_{ij}$ and a $q$-independent piece ${\cal{K}}^{(k)}$ we have,
\begin{align}\label{tofix}
{\cal{L}}^{(k)} &=  {\cal{K}}^{(k)} + \mathcal{Q}^{(k)} + \sum_{4i+6j = 2k} d_{ij} E_4^i E_6^j. 
\end{align}  
To fix the set $d_{\{ij\}}$ we take as input the $q$-expansion coefficients $\{ a_{m}^{(k)}\}_{m=1}^{d_k}$ and fix these by expanding \eqref{tofix} in a Fourier series up to  $\CO \left( q^{d_k }\right) $. This gives $d_k$ equations to fix the $d_k$ unknowns $d_{\{ij\}}$. Let the solutions to these equations be denoted $d_{\{ij\}}^s$.  

The $q$-independent piece at this order, ${\cal{K}}^{(k)}$ is determined by demanding that the $\CO\left(q^0\right)$ term in the Fourier expansion vanishes, which gives,
\begin{align}
{\cal{K}}^{(k)} &= - \lim_{q\rightarrow 0} \left( \mathcal{Q}^{(k)} + \sum_{4i+6j = 2k} d_{ij}^s E_4^i E_6^j \right). 
\end{align}
Therefore using results till previous order, we now can write down, 
\begin{align} {\cal{H}}^{(k)} &=  \exp \left[ F^{(k-1)} +  \frac{{\cal{L}}^{(k)}}{\alpha^{2k}} \right] = \sum_{j\geq0}^{k} \frac{H_j}{h_p^j}, \end{align} 
Finally, note that in order to take the next step in the computation, we need to add $\tilde{h}_k$ to the input data for computation at the next order in the large-$ \alpha $ expansion:
\begin{equation}\label{key}
\left\{ \tilde{h}_0, \tilde{h}_1 ,\dots \tilde{h}_{k-1} \right\} \rightarrow \left\{ \tilde{h}_0, \tilde{h}_1 ,\dots \tilde{h}_{k-1}, \tilde{h}_k = 2^{k+1} k \left( \mathcal{Q}^{(k)} + \sum_{4i+6j =2k} d_{ij}^s E_4^i E_6^j \right) \right\}.
\end{equation}
In this manner, the large-$ h_p $ expansion can be systematically constructed, and it is in this sense that the block is constrained to obey the recursion relation \eqref{eq:MasterRecursion}.

For arriving at the expressions \eqref{F-torus} or \eqref{F-sphere},
if we are to use just the Zamolodchikov $h$-recursions then one also needs to solve for the coefficients of the $E_2^{(k)}$ terms in \eqref{tofix}. Thus the number of unknowns needed to solve for equals the number of integer partitions of $ k $ using only the integers $\lbrace 1, 2, 3 \rbrace $  which we denote $p_{1,2,3}(k)$. This grows quadratically with $k$. In contrast using the additional constraint in form of the recursion \eqref{recrec} we only need to generate $q$-coefficients till order $d_k$ which is linearly bounded. In this sense, the use of the modular anomaly equation allows us to accelerate the computation of the large-$ h_p $ expansion of the block.

\section{Application: heavy-heavy-light OPE coefficients}\label{section:4}
In this section, as an application of the results derived for the Virasoro block, we compute corrections to averaged OPE coefficients. This analysis builds on the work of \cite{Km} and is made possible because we have resummed the $ q $-expansions at each order in the large-$ h_p $ expansion, and consequently have access to the $q\to 1$ (or the high temperature) regime. In what follows, we show how to find the corrections to the averaged heavy-heavy-light OPE coefficient, starting from modular properties of the torus 1-point function. 

The one-point function of a primary $O_h$ on the torus \eqref{tor1}, transforms as a Maass form of weight $(h,\bar{h})$ under modular transformations
\begin{equation}\label{maass}
\vev{O_{h,\bh}}_{\gamma\tau} = (c\tau+d)^h (c\btau+d)^{\hbar} \vev{O_{h,\bh}}_\tau \quad \text{where} \quad \gamma\tau \equiv \frac{a\tau+b}{c\tau+d}
\end{equation}
We also recall that the above quantity can be written as a sum over torus blocks, as follows
\begin{align}\label{tor2}
\vev{O_{h,\bh}}_{\tau,\btau}  = \sum_{p}C_{pOp}\, \cF_{h_p}(q,h,c)\,\Bar\cF_{\bar h_p}(\bq,\bh,c)~. 
\end{align}
The recursive factor for torus block was evaluated perturbatively in large-$h_p$ in \eqref{ht2}. For simplicity, let us just consider just  the first order correction
\begin{align}\label{ht1}
\cF_{h_p}(q,h,c) = \frac{q^{h_p-\frac{c-1}{24}}}{\eta(q)}{\cal H}_{h_p}(q)&= \frac{q^{h_p-\frac{c-1}{24}}}{\eta(q)}\left[ 1  +  \frac{ h(h-1)}{2h_p} \left(\frac{1- E_2(q)}{24} \right) + \cdots\right]. 
\end{align}
We take note of the high-temperature limit  ($q\to 1$ or $\tau\to i0^+$) of the torus block, this will be useful below. Using the S-modular transformations of $\eta(q)$ and $E_2(q)$ we have
\begin{align}\label{hot-block}
\cF_{h_p}(q\to 1,h,c)  
\approx e^{{2\pi i\tau} (h_p-\frac{c}{24})} e^{\frac{\pi i}{12}(\tau+ \frac{1}{\tau})} \sqrt{-i\tau}\left[ 1  -  \frac{ h(h-1)}{48h_p\tau^2}    + \cdots\right]. 
\end{align}
and an analogous relation for the anti-holomorphic block. As we noted above, the higher orders are suppressed in powers of $1/(h_p\tau^2)$. 

Let us now specialize to the case of the rectangular torus, where the modular parameter is $\tau=i\beta/2\pi $. We also take the left- and right-moving temperatures to be independent, $\bar\tau=-i\bar \beta/2\pi$, in order to facilitate the analysis. Equation \eqref{tor2} can be rewritten  as an integral over primaries as follows
\begin{align}\label{int-rep}
\vev{O_{h,\bh}}_{i\beta/L} = \int \mathrm{d}h_p \int \mathrm{d}\bh_p ~ T_{h,\bh}(h_p,\bh_p)\, \cF_{h_p}(e^{-\beta},h,c)\, \Bar\cF_{\bar h_p}(e^{-\bar{\beta}},\bh,c)~,
\end{align}
where, we have introduced the weighted spectral density 
\begin{align}
T_{h,\bh}(h_p,\bh_p) \equiv \sum_p C_{pOp}\, \delta(h-h_p)\,\delta(\bh-\bh_p)~. 
\end{align}
The standard method to obtain high energy asymptotics of the OPE coefficients is via a S-modular transformation for the low temperature result for $\vev{O_{h,\bh}}$, followed by an inverse Laplace transform to extract the weighted spectral density. For $c>1$ theories, there are an infinite number of primaries, growing exponentially at high energies. At high temperatures, the integral \eqref{int-rep} is therefore expected to be dominated by a saddle point of a heavy primary state $(h_p,\bh_p \to \infty)$. This expectation is true for large $c$ theories in which primaries are typical states at high energies, in tune with the weak version of the Eigenstate Thermalization Hypothesis. It is in this regime where we can use our results for the Virasoro blocks in the $1/h_p$ expansion. 

At low temperatures, the expectation value of the primary $O_{h,\bh}$ is dominated by the contribution from the lightest primary $\chi$ which fuses to give $O_{h,\bh}$ in the $\chi\chi$ OPE. Therefore
\begin{align}
\vev{O_{h,\bh}}_{i\beta/L \to \infty} \approx C_{\chi O\chi} \exp\left[- \beta \left(h_\chi - \frac{c}{24}\right)-\bar \beta \left( \bh_\chi - \frac{c}{24}\right)\right]~. 
\end{align}
Upon using the modular property \eqref{maass} for the S-modular transformation, we have the high temperature version to be
\begin{align}
\vev{O_{h,\bh}}_{i\beta/L \to 0} \approx C_{\chi O\chi} \left(\frac{i \beta}{2\pi}\right)^{-h}\left(-\frac{i \bar \beta}{2\pi}\right)^{-\bar h}  \exp\left[-\frac{4\pi^2 }{\beta}\left(h_\chi   - \frac{c}{24}\right)-\frac{4\pi^2}{\bar\beta}\left(\bh_\chi - \frac{c}{24}\right)\right]~. 
\end{align}
 When the $1/h_p$ correction to the high temperature block \eqref{hot-block} is ignored we obtain the result for the leading weighted spectral density, denoted by $T^{(0)}_{h,\bar{h}} ( h_p, \bar{h}_p)$. This is a slight variant of the result of \cite{Km}, since here we do not sum over spins. The inverse Laplace transform leads to the following factorized version $T^{(0)}_{h,\bh}(h_p,\bh_p)$$=C_{\chi O\chi} t_{h}^{(0)}( h_p)\bar t^{(0)}_{\bar h}( \bar h_p)$, analogous to the Cardy formula \cite{Loran:2010bd, Hartman:2014oaa}, with
 \begin{align}\label{invlapla}
 t^{(0)}_h (h_p) &= i^{-h} \oint \mathrm{d}\beta\,\, \left( \frac{\beta}{2\pi} \right)^{ -\tfrac{1}{2} - h } \exp \left[ - \frac{4\pi^2}{\beta} \left( h_\chi - \frac{\hat{c}}{24} \right) + \beta \left( h_p - \frac{\hat{c}}{24} \right)\right] \nn , \\
&\approx i^{-h} \left( \tfrac{\hat{c}}{24} - h_\chi \right)^{-\tfrac{1}{4}-\tfrac{h}{2}} \,\, \bigg( h_p - \tfrac{\hat{c}}{24} \bigg)^{\tfrac{h }{2} + \tfrac{1}{4} } \,\, \exp\left[ 4\pi \sqrt{ \bigg( \tfrac{\hat{c}}{24} - h_\chi \bigg) \bigg(h_p- \tfrac{  \hat{c}}{24} \bigg) } \right], 
\end{align}
and $\bar{t}^{(0)}_{\bh}(\bh_p)$ is given the same formula as the above but with the replacements $i \mapsto -i$, $h \mapsto \bh$ and $h_\chi \mapsto \bh_\chi$. In the above expression, we use the notation $\hat c \equiv c-1$. 
While arriving at the above expression, the Laplace transform was evaluated in the saddle point approximation. The location of the saddle is
\begin{align}
\frac{\beta_* }{2\pi } &= \sqrt{ \frac{ \tfrac{\hat{c}}{24} - h_\chi }{ h_p - \tfrac{\hat{c}}{24 } } } +  \frac{ h  + \tfrac{1}{2}}{ 4\pi ( h_p - \tfrac{\hat{c}}{24 } )} + O\left( 1/h_p^{3/2} \right).
\end{align}
The saddle for $\bar\beta_*$ is takes the same form as the above with the appropriate replacements. Note that the saddle-point approximation is valid when
\begin{align}\label{KM-regime}
\frac{\beta_*^2}{4\pi^2} \left(h_p - \frac{\hat{c}}{24 }\right) \approx \frac{\hat c}{24} - h_\chi \gg 1 \ .
\end{align} 
That is, we require the central charge, $c$, to be large and $\chi$ to be a light operator with $h_\chi \ll \frac{c}{24}$. 
To keep track of the correction to the weighted spectral density arising from $1/h_p$ corrections to the block, we write $T_{h,\bh}(h_p,\bh_p)$ as $T^{(0)}_{h,\bh}(h_p,\bh_p)[1+\delta T_{h,\bh}(h_p,\bh_p) ]$. Here the prefactor $T^{(0)}_{h,\bh}(h_p,\bh_p)$ is the leading result.
Keeping in mind that $T^{(0)}_{h,\bh}(h_p,\bh_p)$ comes from the leading piece of the block without $1/h_p$ corrections, we then have the following equality as a consistency condition
\begin{align}
\int \mathrm{d}h_p \int \mathrm{d}\bar{h}_p \,\,  T ^{(0)}_{h,\bh}(h_p,\bh_p) &\left(  \frac{h  ( h -1 ) \pi^2}{12 h_p \beta^2 }  +\frac{\bar{h}  ( \bar{h} -1 )\pi^2}{12 \bar{h}_p \bar{\beta}^2 } + \cdots \right)   
e^{- \beta( h_p - \tfrac{\hat{c}}{24} ) } e^{-\bar \beta  ( \bar h_p - \tfrac{\hat{c}}{24} ) }
 \\ &= -  \int \mathrm{d}h_p \int \mathrm{d}\bar{h}_p~ T ^{(0)}_{h,\bh}(h_p,\bh_p) \delta T_{h,\bh}(h_p,\bh_p)  ~  e^{-\beta ( h_p - \tfrac{\hat{c}}{24} ) } e^{-\bar \beta( \bar h_p - \tfrac{\hat{c}}{24} ) }. \label{toplug}\nn
\end{align}
where we have used the high temperature limit of the blocks \eqref{hot-block} in \eqref{int-rep}. Writing the first order correction to the spectral density as $\delta T^{(1)}_{h,\bh}(h_p,\bh_p) $ $= \delta  t^{(1)}_{h}(h_p)\delta  t^{(1)}_{\bh}(\bh_p)$ and evaluating the integrals on the LHS, we have
\begin{align}
&\int \mathrm{d}h_p\, t ^{(0)}_{h}(h_p) \delta t ^{(1)}_{h}(h_p) e^{- \beta ( h_p - \tfrac{\hat{c}}{24} ) }  \int \mathrm{d}\bar{h}_p\, \bar{t} ^{(0)}_{\bh}(\bh_p) \delta \bar{t} ^{(1)}_{\bh}(\bh_p) e^{-\bar \beta ( \bar h_p - \tfrac{\hat{c}}{24} ) } \nn \\  
\approx&- i^{-s}\left[\frac{h  ( h -1 )}{48(\tfrac{\hat{c}}{24} - h_\chi) }  + \frac{\bar{h}  ( \bar{h} -1 )}{48(\tfrac{\hat{c}}{24} - \bar{h}_\chi) }\right]
e^{\tfrac{4 \pi^2 }{\bar{\beta}} ( \bar{h}_\chi - \tfrac{\hat{c}}{24 } ) } \left(\frac{\bar{\beta}}{2\pi}\right)^{ - \bar{h} - \tfrac{1}{2}}e^{\tfrac{4 \pi^2  }{\beta} ( h_\chi - \tfrac{\hat{c}}{24 } ) } \left(\frac{{\beta}}{2\pi}\right)^{ - {h} - \tfrac{1}{2} }
\end{align} 
Finally, the product $\delta  t^{(1)}_{h}(h_p)\delta  t^{(1)}_{\bh}(\bh_p)$ can be extracted by an inverse Laplace transform with respect to $\beta$ and $\bar\beta$. The integrals are exactly same as in eq \eqref{invlapla}. This leads to the result
\begin{align}
\delta T^{(1)}_{h,\bh}(h_p,\bh_p)= \delta  t^{(1)}_{h}(h_p)\delta  t^{(1)}_{\bh}(\bh_p)\approx -\left[\frac{h  ( h -1 )}{48(\tfrac{\hat{c}}{24} - h_\chi) }  + \frac{\bar{h}  ( \bar{h} -1 )}{48(\tfrac{\hat{c}}{24} - \bar{h}_\chi) }\right]~.
\end{align}
This shows that the $1/h_p$ corrections to block translate into $1/c$ corrections to the weighted spectral density, by virtue of saddle point approximation \eqref{KM-regime}. The averaged heavy-heavy-light coefficient can be obtained by dividing out the weighted spectral density by the density of states. The final result is 
\begin{align}
\overline{C_{HOH}} \approx \overline{C^{(0)}_{HOH}} \left[1-\frac{h  ( h -1 )}{48(\tfrac{\hat{c}}{24} - h_\chi) }  -\frac{\bar{h}  ( \bar{h} -1 )}{48(\tfrac{\hat{c}}{24} - \bar{h}_\chi) }+\cdots\right]~. 
\end{align}
where
\begin{align}
\overline{C^{(0)}_{HOH}} &= \frac{t^{(0)}_h(h_p)}{\rho(h_p)}\frac{\bar t^{(0)}_{\bh}(\bh_p)}{\rho(\bh_p)} \nn \\
&\approx  \frac{C_{\chi O \chi} i^{-s} }{2\pi^2 } \frac{ \left( h_p - \tfrac{\hat{c}}{24} \right)^{\tfrac{h+1}{2} }}{ \left( \tfrac{\hat{c}}{24} - h_\chi \right)^{\tfrac{1}{4} + \tfrac{h}{2} } }  \frac{ \left( \bar h_p - \tfrac{\hat{c}}{24} \right)^{\tfrac{\bar h+1}{2} }}{ \left( \tfrac{\hat{c}}{24} -\bar  h_\chi \right)^{\tfrac{1}{4} + \tfrac{\bar h}{2} } } \exp \left[ - \frac{\pi \hat{c}}{6}  \left( 1 - \sqrt{ 1 - \tfrac{24 h_\chi}{\hat{c}} } \right) \sqrt{ \tfrac{24 h_p}{\hat{c}}-1} \right] \nn\\
&\qquad\times  \exp \left[ - \frac{\pi \hat{c}}{6}  \left( 1 - \sqrt{ 1 - \tfrac{24 \bar h_\chi}{\hat{c}} } \right) \sqrt{ \tfrac{24 \bar h_p}{\hat{c}}-1} \right] 
\end{align}
The corrections arising from higher orders in $1/h_p$ can also be systematically worked out in much the same way. As we have mentioned, the $1/h_p$ expansion of the block corresponds to a $1/c$ expansion for the averaged OPE coefficient. 
A similar analysis can also be carried out for the averaged light-light-heavy OPE coefficient using the crossing symmetry of the 4-point correlator on the plane in the pillow frame. 
\section{Conclusions}\label{section:5}


Virasoro blocks lie at the heart of two-dimensional conformal field theories. Except in a handful of cases, the blocks are not known in closed form. This has motivated numerous attempts to understand them in various regimes with the hope that simplifications will arise. Such studies  has often resulted in insightful revelations about their structure.

In this work we have studied the $ 1 $-point block on the torus and the $ 4 $-point block on the sphere in the regime of heavy intermediate exchange.  Our analysis shows that each order in the large-$h_p$ expansion can be resummed into polynomials of the Eisenstein series. Although a partial explanation of the appearance of quasimodular forms was provided from purely CFT methods via the fusion kernel, a clearer understanding of this arose from the $ 2 $d/$ 4 $d correspondence. We established that a modular anomaly equation constrains the block, and further that it may be used to constructively build up the Virasoro blocks non-perturbatively (in $\tau$) at every order in the $1/h_p$ expansion. Moreover, it was noted that the modular anomaly constraint in the block is compactly encoded in the KPZ equation \eqref{KPZ} and its solution is directly related to the block \eqref{HF-relation}.  We hope that connecting these dots might allow us to pin down the CFT origin of the modular anomaly in the blocks. 

It is curious that the recursion relations are of the same form for both the $ 1 $-point block on the torus and the $ 4 $-point block on the sphere. This resemblance deserves further attention. In fact, there are identities due to Poghossian relating these two blocks \cite{Poghossian:2009mk}. However, a direct translation of these identities to a correspondence between the recursion relations is not known. Understanding this would shed more light on the structure of the blocks. 

The closed forms (in cross-ratio) at each order in the large-$h_p$ expansion for the $ 4 $-point block on the sphere can provide a new window into studying the block in a Lorentzian setup and allow one to focus on  scrambling/late time behaviour. Such studies have appeared in the past in the context of the out-of-time-ordered correlators \cite{Liu:2018iki, Hampapura:2018otw,Chang:2018nzm}, toy versions of the black hole information puzzle in AdS$_3$/CFT$_2$ \cite{Chen:2017yze}, relations to Wilson lines in AdS$_3$ and OPE inversion \cite{Kraus:2018zrn} and also entanglement entropy in quantum quenches \cite{Kusuki:2018wpa}. It would be worthwhile to apply the results of the blocks derived here to find refinements to those investigations.

Virasoro blocks for higher-point correlation functions and those on higher genus Riemann surfaces have more parameters/moduli as well as more intermediate channels. Based on the strategies by Zamolodchikov, a general recursive representations for the blocks corresponding to higher-point correlation functions on higher genus Riemann surfaces have also been developed \cite{xy}. It would be worthwhile to investigate the regime of heavy exchanges for these blocks and study the (quasi-)modular structures that appear. The $ 4 $-point block treated here can also be considered in the more general case of unequal external operator dimensions. It should be straightforward to show, for example, that Jacobi theta functions will appear in the large-$h_p$ expansion in such cases, as the $ 2 $d/$ 4 $d correspondence would lead one to believe.

There also exist extensions of the 2d/4d story for CFTs with higher spin symmetries \cite{Wyllard:2009hg}. It is reasonable to expect that analogous modular features will appear for $\mathcal{W}_N$ conformal blocks in the heavy exchange expansion. One can also hope that super-Virasoro blocks have similar structures that can be uncovered.

An analysis complementary to that considered here would be to consider the Virasoro blocks in the large-$c$ expansion instead of the large-$h_p$ one. Although modular features are not manifest, the first few orders in the large-$c$ expansion can be written by  using a combination of hypergeometric functions \cite{Beccaria:2015shq,Bombini:2018jrg,Fitzpatrick:2016thx}. It would be interesting to find an analogue of the modular anomaly recursion for the terms in this expansion. However, it is far from clear how such a mechanism would work in the absence of some additional constraints. On the other hand, the block is known to exponentiate in the $c\to\infty$ regime. This might offer some crucial hints on how to proceed.

\section*{Acknowledgements}
We thank Sujay Ashok, Carlos Cardona, Per Kraus, Wolfgang Lerche, Guglielmo Lockhart, Jaewon Song and Alexander Zhiboedov for discussions.
We are grateful to the authors of \cite{Chen:2017yze} for their publicly available code for 4-point Virasoro blocks which was used in this work. 
DD would like to acknowledge the support provided by the Max Planck Partner Group grant MAXPLA/PHY/2018577. MR  acknowledges  support  from  the  Infosys  Endowment  for Research into the Quantum Structure of Spacetime.
\appendix

\section*{Appendix}

\section{Eisenstein series}

The Eisenstein series were used extensively in the main text. In this appendix we provide their definitions (which fixes our normalization conventions) and list few of their properties. 

The $q$-expansions  for the Eisenstein series are
as follows ($q=e^{2\pi i \tau}$)
\begin{equation} \label{Ek-def}
E_{2k}(q) = 1- \frac{4k}{B_{2k}}\sum_{n=1}^\infty \sigma_{2k-1}(n)q^n = 1+ \frac{2}{\zeta(1-2k)}\sum_{n=1}^\infty \frac{n^{2k-1}q^n}{1-q^n}. 
\end{equation}
Here, $\sigma_n$ is the divisor function, $B_{m}$ are the Bernoulli numbers and $\zeta(p)$ is the Riemann-zeta function. Another representation of the Eisenstein series in terms of the lattice sums
\begin{align}\label{lattice-Ek}
E_{2k}(\tau) = \frac{1}{2\zeta(2k)} \sum_{(m,n)\in \mathbb{Z}^2\backslash (0,0)} \frac{1}{(m+n\tau)^{2k}}. 
\end{align}
Upon modular transformations, SL$(2,\mathbb{Z})$, the Eisenstein series transform as 
\begin{align}\label{mod-prop-Ek}
E_2\left(\frac{a\tau+b}{c\tau+d}\right)= (c\tau+d)^2 E_2(\tau) + \frac{6c}{i\pi}(c\tau+d)\, , \quad E_{2k}\left(\frac{a\tau+b}{c\tau+d}\right) = (c\tau+d)^{2k} E_{2k}(\tau) \text{ for } k\geq2 \, .
\end{align}
That is, $E_2$ is quasimodular while the other Eisenstein series are modular forms. The modular forms $E_4$ and $E_6$ generate the ring of modular forms of any even weight. This implies that $E_{2k\geq8}$ can be written in terms of polynomials of $E_4$ and $E_6$. This fact plays a role in writing higher orders in the $1/h_p$ expansion in terms of $E_2$, $E_4$ and $E_6$ alone. The plots of the arguments of $E_{2,4,6}(q)$ on the unit-disk on the $q$-plane are shown below. 
\begin{figure}[!h]
	\centering
	\includegraphics[width=\linewidth]{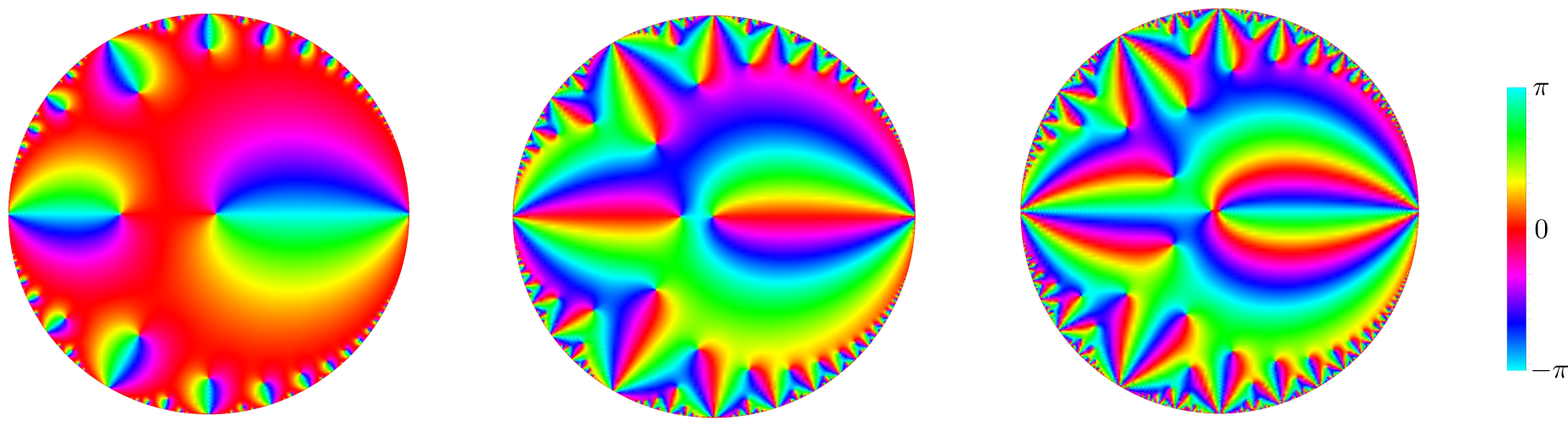}
	\caption{\textbf{$\arg[E_{2k}(q)]$  on the unit disk: } for [Left] $E_2(q)$, [Centre] $E_4(q)$ and [Right] $E_6(q)$. The real and imaginary parts can also be plotted and they show similar structures. Details regarding generating these plots can be found in \cite{stackexchange}. }
	\label{fig:Ediscs}
\end{figure}

\section{Higher orders in the $1/h_p$ expansion}
\label{appB}
\subsection*{Torus 1-point block}
The 3rd order in the $1/h_p$ expansion of $\cH_{h_p}(h,q)$ appearing in the 1-point torus block is the following
\begin{footnotesize}
	\begin{align}
	&\frac{h(h-1) }{23224320h_p^3}\bigg[1048 c^2+c (6 h (77 h+339)-6320)+(h-2) (h (7 h (5 h+72)+1481)-6236)\nn \\
	&+105 E_2^2 (h-3) (h-2) (4 c+(h-1) h-4)-21 E_2 \left((42 c-97) h^2+6 (13-7 c) h+40 (c-1)^2+5 h^4+14 h^3\right)\nn \\
	&-42 E_4 (c+2 h-4) (4 c+(h-1) h-4)-35 E_2^3 (h-5) (h-4) (h-3) (h-2)\nn \\
	&+42 E_2 E_4 (h-5) (h-4) (c+2 h-4)-8 E_6 \left(5 c^2+c (18 h-55)+11 (h-5) (h-2)\right) \bigg]~.
	\end{align}
\end{footnotesize}\noindent
The 4th order is 
\begin{footnotesize}
	\begin{align}
	&\frac{h(h-1) }{22295347200h_p^4}\bigg[50160 c^3+4 c^2 (h (7781 h+41899)-136620)+4 c (h (h (3 h (385 h+5238)+28133)-438122)+663300)\nn \\ \nn &+(h-2) (h (h (h (35 h (5 h+139)+43291)+35647)-1286498)+2348520)\nn \\
	&+20 E_2 (-1680 c^3+4 c^2 (1260-703 (h-1) h)-16 c ((h-1) h (3 h (14 h+59)-710)+315)\nn \\&\qquad\qquad\qquad-(h-2) (h (h (h (7 h (5 h+67)+767)-8725)+8294)+840)) \nn \\
	&+210 E_2^2 (h-3) (h-2) \left(120 c^2+c (62 (h-1) h-240)+(h-1) h (h (5 h+19)-98)+120\right)\nn \\
	&-700 E_2^3 (h-5) (h-4) (h-3) (h-2) (6 c+(h-3) (h+2))\nn \\
	&+175 E_2^4 (h-7) (h-6) (h-5) (h-4) (h-3) (h-2)\nn \\
	&-84 E_4 (c+2 h-4) \left(120 c^2+c (62 (h-1) h-240)+(h-1) h (h (5 h+19)-98)+120\right) \nn \\
	&+840 E_2 E_4 (h-5) (h-4) (c+2 h-4) (6 c+(h-3) (h+2))\nn 
	\end{align}
	\begin{align}
	&-420 E_2^2 E_4 (h-7) (h-6) (h-5) (h-4) (c+2 h-4)\nn \\
	&+12 E_4^2  (-140 c^3+c^2 (h (7 h-887)+3500)+4 c (h (h (7 h-421)+3134)-5635)+4 (h-7) (h-5) (h-2) (7 h-127) )\nn \\
	&+160 E_2 E_6 (h-7) (h-6) \left(5 c^2+c (18 h-55)+11 (h-5) (h-2)\right)\nn \\&-160 E_6 (6 c+(h-3) (h+2)) \left(5 c^2+c (18 h-55)+11 (h-5) (h-2)\right)  \bigg]~. 
	\end{align}
\end{footnotesize}\noindent
In addition to vanishing for $h=0,1$, the above expressions also vanish for $c=0,h=2$ and $c=-2,h=3$ \cite{Beccaria:2016nnb}. 

\subsection*{Sphere 4-point block}
The 3rd order in the $1/h_p$ expansion of $\cH(c,h,h_p,q)$ appearing in the 4-point sphere block \eqref{Vir-block} is the following
\begin{footnotesize}
	\begin{align}
	&\frac{1}{11890851840h_p^3}\bigg[35 c^6+42 c^5 (79-160 h)+c^4 (1344 h (400 h-283)+56017)\nn \\
	&-4 c^3 (224 h (32 h (800 h-513)+647)+240167)+c^2 (128 h (16 h (1344 h (200 h-59)-72013)+461749)+2664021)\nn \\
	&+c (29660102-64 h (64 h (96 h (112 h (160 h+53)-9861)-3595)+7556053))\nn  \\
	&+64 h (16 h (128 h (8 h (448 h (80 h+99)-12207)-225177)+13419349)-20775123)+31536855
	\nn \\
	&+105 E_2^2 (c-32 h+1) (c-32 h+5) (c-32 h+9) (c-32 h+13)  (c^2-64 (c+3) h+38 c+1024 h^2-27 )
	\nn \\
	&
	-336 E_4  (c^2-64 (c+3) h+38 c+1024 h^2-27 )  (1024 (c-41) h^2+32 c (58-5 c) h+c (c (3 c-17)-111)\nn\\&\qquad\qquad+32768 h^3+4576 h-115 )\nn \\
	&+21 E_2 (c-32 h+1) (c-32 h+5)  (-5 c^4+4 c^3 (160 h-87)-6 c^2 (320 h (16 h-11)+513)\nn\\&\qquad\qquad+4 c (32 h (32 h (160 h-69)-317)+2669)-128 h (16 h (64 h (40 h+9)-893)+1917)+3555 )\nn \\
	&-128 E_6  (187 c^4-4 c^3 (2016 h+919)+c^2 (128 (1985-368 h) h-982)\nn\\&\qquad\qquad+4 c (32 h (32 h (800 h-733)-5693)+11749)+128 h (16 h (64 h (88 h-333)+13277)-20037)+63315 )\nn \\
	&+336 E_2 E_4 (c-32 h+17) (c-32 h+21) \nn \\
	&\qquad\qquad \times (1024 (c-41) h^2+32 c (58-5 c) h+c (c (3 c-17)-111)+32768 h^3+4576 h-115 )\nn \\
	&-35 E_2^3 (c-32 h+1) (c-32 h+5) (c-32 h+9) (c-32 h+13) (c-32 h+17) (c-32 h+21) \bigg].
	\end{align}
\end{footnotesize}\noindent
The above expression can be seen to vanish for $c=1,h=1/16$ and $c=25,h=15/16$.

\bibliography{refsBlocks2020}

\providecommand{\href}[2]{#2}\begingroup\begin{thebibliography}{10}

\bibitem{zamu1}
A.~B. Zamolodchikov, {\it Conformal symmetry in two-dimensional space:
  Recursion representation of conformal block},
  \href{http://dx.doi.org/10.1007/BF01022967}{{\sf Theoretical and Mathematical
  Physics} {\sf {73} }{\sf no.~1, }{\sf (Oct, 1987) }{\sf 1088--1093}}.

\bibitem{dattaper}
M.~Besken, S.~Datta, and P.~Kraus, {\it {Semi-classical Virasoro blocks: proof
  of exponentiation}},  \href{http://dx.doi.org/10.1007/JHEP01(2020)109}{{\sf
  JHEP} {\sf {01} }{\sf (2020) }{\sf 109}},
  \href{http://arxiv.org/abs/1910.04169}{{\ttfamily arXiv:1910.04169
  [hep-th]}}.

\bibitem{zamolodchikov1986two}
A.~B. Zamolodchikov, {\it Two-dimensional conformal symmetry and critical
  four-spin correlation functions in the ashkin-teller model},  {\sf Sov.
  Phys.-JETP} {\sf {63} }{\sf (1986) }{\sf 1061--1066}.

\bibitem{fkw}
A.~Fitzpatrick, J.~Kaplan, and M.~T. Walters, {\it {Universality of
  Long-Distance AdS Physics from the CFT Bootstrap}},
  \href{http://dx.doi.org/10.1007/JHEP08(2014)145}{{\sf JHEP} {\sf {08} }{\sf
  (2014) }{\sf 145}}, \href{http://arxiv.org/abs/1403.6829}{{\ttfamily
  arXiv:1403.6829 [hep-th]}}.

\bibitem{harlow}
D.~Harlow, J.~Maltz, and E.~Witten, {\it {Analytic Continuation of Liouville
  Theory}},  \href{http://dx.doi.org/10.1007/JHEP12(2011)071}{{\sf JHEP} {\sf
  {12} }{\sf (2011) }{\sf 071}},
  \href{http://arxiv.org/abs/1108.4417}{{\ttfamily arXiv:1108.4417 [hep-th]}}.

\bibitem{zamu2}
A.~B. Zamolodchikov, {\it {Conformal symmetry in two-dimensions : An explicit
  recurrence formula for the conformal partial wave amplitude}},
\href{http://dx.doi.org/10.1007/BF01214585}{{\sf Commun. Math. Phys.} {\sf {96}
  }{\sf (1984) }{\sf 419--422}}.

\bibitem{Hadasz:2009db}
L.~Hadasz, Z.~Jaskolski, and P.~Suchanek, {\it {Recursive representation of the
  torus 1-point conformal block}},
  \href{http://dx.doi.org/10.1007/JHEP01(2010)063}{{\sf JHEP} {\sf {01} }{\sf
  (2010) }{\sf 063}}, \href{http://arxiv.org/abs/0911.2353}{{\ttfamily
  arXiv:0911.2353 [hep-th]}}.

\bibitem{xy}
M.~Cho, S.~Collier, and X.~Yin, {\it {Recursive Representations of Arbitrary
  Virasoro Conformal Blocks}},
  \href{http://dx.doi.org/10.1007/JHEP04(2019)018}{{\sf JHEP} {\sf {04} }{\sf
  (2019) }{\sf 018}}, \href{http://arxiv.org/abs/1703.09805}{{\ttfamily
  arXiv:1703.09805 [hep-th]}}.

\bibitem{Km}
P.~Kraus and A.~Maloney, {\it {A cardy formula for three-point coefficients or
  how the black hole got its spots}},
  \href{http://dx.doi.org/10.1007/JHEP05(2017)160}{{\sf JHEP} {\sf {05} }{\sf
  (2017) }{\sf 160}}, \href{http://arxiv.org/abs/1608.03284}{{\ttfamily
  arXiv:1608.03284 [hep-th]}}.

\bibitem{modcharged}
D.~Das, S.~Datta, and S.~Pal, {\it {Charged structure constants from
  modularity}},  \href{http://dx.doi.org/10.1007/JHEP11(2017)183}{{\sf JHEP}
  {\sf {11} }{\sf (2017) }{\sf 183}},
  \href{http://arxiv.org/abs/1706.04612}{{\ttfamily arXiv:1706.04612
  [hep-th]}}.

\bibitem{modoff}
E.~M. Brehm, D.~Das, and S.~Datta, {\it {Probing thermality beyond the
  diagonal}},  \href{http://dx.doi.org/10.1103/PhysRevD.98.126015}{{\sf Phys.
  Rev. D} {\sf {98} }{\sf no.~12, }{\sf (2018) }{\sf 126015}},
  \href{http://arxiv.org/abs/1804.07924}{{\ttfamily arXiv:1804.07924
  [hep-th]}}.

\bibitem{modoff2}
A.~Romero-Berm{\'u}dez, P.~Sabella-Garnier, and K.~Schalm, {\it {A Cardy
  formula for off-diagonal three-point coefficients; or, how the geometry
  behind the horizon gets disentangled}},
  \href{http://dx.doi.org/10.1007/JHEP09(2018)005}{{\sf JHEP} {\sf {09} }{\sf
  (2018) }{\sf 005}}, \href{http://arxiv.org/abs/1804.08899}{{\ttfamily
  arXiv:1804.08899 [hep-th]}}.

\bibitem{modoff3}
Y.~Hikida, Y.~Kusuki, and T.~Takayanagi, {\it {Eigenstate thermalization
  hypothesis and modular invariance of two-dimensional conformal field
  theories}},  \href{http://dx.doi.org/10.1103/PhysRevD.98.026003}{{\sf Phys.
  Rev. D} {\sf {98} }{\sf no.~2, }{\sf (2018) }{\sf 026003}},
  \href{http://arxiv.org/abs/1804.09658}{{\ttfamily arXiv:1804.09658
  [hep-th]}}.

\bibitem{modpillow}
D.~Das, S.~Datta, and S.~Pal, {\it {Universal asymptotics of three-point
  coefficients from elliptic representation of Virasoro blocks}},
  \href{http://dx.doi.org/10.1103/PhysRevD.98.101901}{{\sf Phys. Rev. D} {\sf
  {98} }{\sf no.~10, }{\sf (2018) }{\sf 101901}},
  \href{http://arxiv.org/abs/1712.01842}{{\ttfamily arXiv:1712.01842
  [hep-th]}}.

\bibitem{Ponsot:1999uf}
B.~Ponsot and J.~Teschner, {\it {Liouville bootstrap via harmonic analysis on a
  noncompact quantum group}},
  \href{http://arxiv.org/abs/hep-th/9911110}{{\ttfamily arXiv:hep-th/9911110}}.

\bibitem{Ponsot:2000mt}
B.~Ponsot and J.~Teschner, {\it {Clebsch-Gordan and Racah-Wigner coefficients
  for a continuous series of representations of U(q)(sl(2,R))}},
  \href{http://dx.doi.org/10.1007/PL00005590}{{\sf Commun. Math. Phys.} {\sf
  {224} }{\sf (2001) }{\sf 613--655}},
  \href{http://arxiv.org/abs/math/0007097}{{\ttfamily arXiv:math/0007097}}.

\bibitem{collier-mod}
S.~Collier, Y.~Gobeil, H.~Maxfield, and E.~Perlmutter, {\it {Quantum Regge
  Trajectories and the Virasoro Analytic Bootstrap}},
  \href{http://dx.doi.org/10.1007/JHEP05(2019)212}{{\sf JHEP} {\sf {05} }{\sf
  (2019) }{\sf 212}}, \href{http://arxiv.org/abs/1811.05710}{{\ttfamily
  arXiv:1811.05710 [hep-th]}}.

\bibitem{brehmdas}
E.~M. Brehm and D.~Das, {\it {Aspects of the S transformation Bootstrap}},
  \href{http://dx.doi.org/10.1088/1742-5468/ab7f36}{{\sf J. Stat. Mech.} {\sf
  {2005} }{\sf (2020) }{\sf 053103}},
  \href{http://arxiv.org/abs/1911.02309}{{\ttfamily arXiv:1911.02309
  [hep-th]}}.

\bibitem{maloney-mod}
S.~Collier, A.~Maloney, H.~Maxfield, and I.~Tsiares, {\it {Universal Dynamics
  of Heavy Operators in CFT$_2$}},
  \href{http://arxiv.org/abs/1912.00222}{{\ttfamily arXiv:1912.00222
  [hep-th]}}.

\bibitem{Cheng:2020srs}
M.~C. Cheng, T.~Gannon, and G.~Lockhart, {\it {Modular Exercises for Four-Point
  Blocks -- I}},  \href{http://arxiv.org/abs/2002.11125}{{\ttfamily
  arXiv:2002.11125 [hep-th]}}.

\bibitem{Cardona:2020cfy}
C.~Cardona, {\it {Virasoro blocks at large exchange dimension}},
  \href{http://arxiv.org/abs/2006.01237}{{\ttfamily arXiv:2006.01237
  [hep-th]}}.

\bibitem{KashaniPoor:2012wb}
A.-K. Kashani-Poor and J.~Troost, {\it {The toroidal block and the genus
  expansion}},  \href{http://dx.doi.org/10.1007/JHEP03(2013)133}{{\sf JHEP}
  {\sf {03} }{\sf (2013) }{\sf 133}},
  \href{http://arxiv.org/abs/1212.0722}{{\ttfamily arXiv:1212.0722 [hep-th]}}.

\bibitem{Kashani-Poor:2013oza}
A.-K. Kashani-Poor and J.~Troost, {\it {Transformations of Spherical Blocks}},
  \href{http://dx.doi.org/10.1007/JHEP10(2013)009}{{\sf JHEP} {\sf {10} }{\sf
  (2013) }{\sf 009}}, \href{http://arxiv.org/abs/1305.7408}{{\ttfamily
  arXiv:1305.7408 [hep-th]}}.

\bibitem{Kashani-Poor:2014mua}
A.-K. Kashani-Poor and J.~Troost, {\it {Quantum geometry from the toroidal
  block}},  \href{http://dx.doi.org/10.1007/JHEP08(2014)117}{{\sf JHEP} {\sf
  {08} }{\sf (2014) }{\sf 117}},
  \href{http://arxiv.org/abs/1404.7378}{{\ttfamily arXiv:1404.7378 [hep-th]}}.

\bibitem{agt}
L.~F. Alday, D.~Gaiotto, and Y.~Tachikawa, {\it {Liouville Correlation
  Functions from Four-dimensional Gauge Theories}},
  \href{http://dx.doi.org/10.1007/s11005-010-0369-5}{{\sf Lett. Math. Phys.}
  {\sf {91} }{\sf (2010) }{\sf 167--197}},
  \href{http://arxiv.org/abs/0906.3219}{{\ttfamily arXiv:0906.3219 [hep-th]}}.

\bibitem{Pestun:2016zxk}
V.~Pestun {\em et al.}, {\it {Localization techniques in quantum field
  theories}},  \href{http://dx.doi.org/10.1088/1751-8121/aa63c1}{{\sf J. Phys.
  A} {\sf {50} }{\sf no.~44, }{\sf (2017) }{\sf 440301}},
  \href{http://arxiv.org/abs/1608.02952}{{\ttfamily arXiv:1608.02952
  [hep-th]}}.

\bibitem{Minahan:1997if}
J.~Minahan, D.~Nemeschansky, and N.~Warner, {\it {Instanton expansions for mass
  deformed N=4 superYang-Mills theories}},
  \href{http://dx.doi.org/10.1016/S0550-3213(98)00314-9}{{\sf Nucl. Phys. B}
  {\sf {528} }{\sf (1998) }{\sf 109--132}},
  \href{http://arxiv.org/abs/hep-th/9710146}{{\ttfamily arXiv:hep-th/9710146}}.

\bibitem{Billo:2015pjb}
M.~Bill{\'o}, M.~Frau, F.~Fucito, A.~Lerda, and J.~Morales, {\it {S-duality and
  the prepotential in $ \mathcal{N}={2}^{\star } $ theories (I): the ADE
  algebras}},  \href{http://dx.doi.org/10.1007/JHEP11(2015)024}{{\sf JHEP} {\sf
  {11} }{\sf (2015) }{\sf 024}},
  \href{http://arxiv.org/abs/1507.07709}{{\ttfamily arXiv:1507.07709
  [hep-th]}}.

\bibitem{Ashok:2016oyh}
S.~Ashok, E.~Dell'Aquila, A.~Lerda, and M.~Raman, {\it {S-duality, triangle
  groups and modular anomalies in $ \mathcal{N}=2 $ SQCD}},
  \href{http://dx.doi.org/10.1007/JHEP04(2016)118}{{\sf JHEP} {\sf {04} }{\sf
  (2016) }{\sf 118}}, \href{http://arxiv.org/abs/1601.01827}{{\ttfamily
  arXiv:1601.01827 [hep-th]}}.

\bibitem{BCOVReview}
A.~{Kanazawa} and J.~{Zhou}, {\it {Lectures on BCOV holomorphic anomaly
  equations}},  {\sf arXiv e-prints} {\sf (Sept., 2014) }{\sf arXiv:1409.4105},
  \href{http://arxiv.org/abs/1409.4105}{{\ttfamily arXiv:1409.4105 [math.AG]}}.

\bibitem{Fateev:2009aw}
V.~Fateev and A.~Litvinov, {\it {On AGT conjecture}},
  \href{http://dx.doi.org/10.1007/JHEP02(2010)014}{{\sf JHEP} {\sf {02} }{\sf
  (2010) }{\sf 014}}, \href{http://arxiv.org/abs/0912.0504}{{\ttfamily
  arXiv:0912.0504 [hep-th]}}.

\bibitem{Huang:2012kn}
M.-x. Huang, {\it {On Gauge Theory and Topological String in
  Nekrasov-Shatashvili Limit}},
  \href{http://dx.doi.org/10.1007/JHEP06(2012)152}{{\sf JHEP} {\sf {06} }{\sf
  (2012) }{\sf 152}}, \href{http://arxiv.org/abs/1205.3652}{{\ttfamily
  arXiv:1205.3652 [hep-th]}}.

\bibitem{1302}
M.~Billo, M.~Frau, L.~Gallot, A.~Lerda, and I.~Pesando, {\it {Deformed N=2
  theories, generalized recursion relations and S-duality}},
  \href{http://dx.doi.org/10.1007/JHEP04(2013)039}{{\sf JHEP} {\sf {04} }{\sf
  (2013) }{\sf 039}}, \href{http://arxiv.org/abs/1302.0686}{{\ttfamily
  arXiv:1302.0686 [hep-th]}}.

\bibitem{1307}
M.~Billo, M.~Frau, L.~Gallot, A.~Lerda, and I.~Pesando, {\it {Modular anomaly
  equation, heat kernel and S-duality in $N=2$ theories}},
  \href{http://dx.doi.org/10.1007/JHEP11(2013)123}{{\sf JHEP} {\sf {11} }{\sf
  (2013) }{\sf 123}}, \href{http://arxiv.org/abs/1307.6648}{{\ttfamily
  arXiv:1307.6648 [hep-th]}}.

\bibitem{Beccaria:2016nnb}
M.~Beccaria and G.~Macorini, {\it {Exact partition functions for the
  $\Omega$-deformed $ \mathcal{N}={2}^{\ast } $ SU(2) gauge theory}},
  \href{http://dx.doi.org/10.1007/JHEP07(2016)066}{{\sf JHEP} {\sf {07} }{\sf
  (2016) }{\sf 066}}, \href{http://arxiv.org/abs/1606.00179}{{\ttfamily
  arXiv:1606.00179 [hep-th]}}.

\bibitem{Nemkov}
N.~Nemkov, {\it {On new exact conformal blocks and Nekrasov functions}},
  \href{http://dx.doi.org/10.1007/JHEP12(2016)017}{{\sf JHEP} {\sf {12} }{\sf
  (2016) }{\sf 017}}, \href{http://arxiv.org/abs/1606.05324}{{\ttfamily
  arXiv:1606.05324 [hep-th]}}.

\bibitem{Maldacena:2015iua}
J.~Maldacena, D.~Simmons-Duffin, and A.~Zhiboedov, {\it {Looking for a bulk
  point}},  \href{http://dx.doi.org/10.1007/JHEP01(2017)013}{{\sf JHEP} {\sf
  {01} }{\sf (2017) }{\sf 013}},
  \href{http://arxiv.org/abs/1509.03612}{{\ttfamily arXiv:1509.03612
  [hep-th]}}.

\bibitem{Perlmutter:2015iya}
E.~Perlmutter, {\it {Virasoro conformal blocks in closed form}},
  \href{http://dx.doi.org/10.1007/JHEP08(2015)088}{{\sf JHEP} {\sf {08} }{\sf
  (2015) }{\sf 088}}, \href{http://arxiv.org/abs/1502.07742}{{\ttfamily
  arXiv:1502.07742 [hep-th]}}.

\bibitem{Chen:2017yze}
H.~Chen, C.~Hussong, J.~Kaplan, and D.~Li, {\it {A Numerical Approach to
  Virasoro Blocks and the Information Paradox}},
  \href{http://dx.doi.org/10.1007/JHEP09(2017)102}{{\sf JHEP} {\sf {09} }{\sf
  (2017) }{\sf 102}}, \href{http://arxiv.org/abs/1703.09727}{{\ttfamily
  arXiv:1703.09727 [hep-th]}}.

\bibitem{Kusuki:2018nms}
Y.~Kusuki, {\it {Large $c$ Virasoro Blocks from Monodromy Method beyond Known
  Limits}},  \href{http://dx.doi.org/10.1007/JHEP08(2018)161}{{\sf JHEP} {\sf
  {08} }{\sf (2018) }{\sf 161}},
  \href{http://arxiv.org/abs/1806.04352}{{\ttfamily arXiv:1806.04352
  [hep-th]}}.

\bibitem{Kusuki:2018wcv}
Y.~Kusuki, {\it {New Properties of Large-$c$ Conformal Blocks from Recursion
  Relation}},  \href{http://dx.doi.org/10.1007/JHEP07(2018)010}{{\sf JHEP} {\sf
  {07} }{\sf (2018) }{\sf 010}},
  \href{http://arxiv.org/abs/1804.06171}{{\ttfamily arXiv:1804.06171
  [hep-th]}}.

\bibitem{Besken:2019bsu}
M.~Be\c{s}ken, S.~Datta, and P.~Kraus, {\it {Quantum thermalization and
  Virasoro symmetry}},  \href{http://dx.doi.org/10.1088/1742-5468/ab900b}{{\sf
  J. Stat. Mech.} {\sf {2006} }{\sf (2020) }{\sf 063104}},
  \href{http://arxiv.org/abs/1907.06661}{{\ttfamily arXiv:1907.06661
  [hep-th]}}.

\bibitem{Nemkov2}
N.~Nemkov, {\it {On modular transformations of toric conformal blocks}},
  \href{http://dx.doi.org/10.1007/JHEP10(2015)039}{{\sf JHEP} {\sf {10} }{\sf
  (2015) }{\sf 039}}, \href{http://arxiv.org/abs/1504.04360}{{\ttfamily
  arXiv:1504.04360 [hep-th]}}.

\bibitem{Galakhov:2012gw}
D.~Galakhov, A.~Mironov, and A.~Morozov, {\it {S-duality as a beta-deformed
  Fourier transform}},  \href{http://dx.doi.org/10.1007/JHEP08(2012)067}{{\sf
  JHEP} {\sf {08} }{\sf (2012) }{\sf 067}},
  \href{http://arxiv.org/abs/1205.4998}{{\ttfamily arXiv:1205.4998 [hep-th]}}.

\bibitem{Nemkov1307}
N.~Nemkov, {\it {S-duality as Fourier transform for arbitrary
  $\epsilon_1,\epsilon_2$}},
  \href{http://dx.doi.org/10.1088/1751-8113/47/10/105401}{{\sf J. Phys. A} {\sf
  {47} }{\sf no.~10, }{\sf (2014) }{\sf 105401}},
  \href{http://arxiv.org/abs/1307.0773}{{\ttfamily arXiv:1307.0773 [hep-th]}}.

\bibitem{DLMF}
{\it {NIST Digital Library of Mathematical Functions}},  2020.
\newblock \url{https://dlmf.nist.gov/20.6}.

\bibitem{Iqbal:2007ii}
A.~Iqbal, C.~Kozcaz, and C.~Vafa, {\it {The Refined topological vertex}},
  \href{http://dx.doi.org/10.1088/1126-6708/2009/10/069}{{\sf JHEP} {\sf {10}
  }{\sf (2009) }{\sf 069}},
  \href{http://arxiv.org/abs/hep-th/0701156}{{\ttfamily arXiv:hep-th/0701156}}.

\bibitem{Billo:2015jyt}
M.~Bill{\'o}, M.~Frau, F.~Fucito, A.~Lerda, and J.~Morales, {\it {S-duality and
  the prepotential of $ \mathcal{N}={2}^{\star } $ theories (II): the
  non-simply laced algebras}},
  \href{http://dx.doi.org/10.1007/JHEP11(2015)026}{{\sf JHEP} {\sf {11} }{\sf
  (2015) }{\sf 026}}, \href{http://arxiv.org/abs/1507.08027}{{\ttfamily
  arXiv:1507.08027 [hep-th]}}.

\bibitem{Bershadsky:1993ta}
M.~Bershadsky, S.~Cecotti, H.~Ooguri, and C.~Vafa, {\it {Holomorphic anomalies
  in topological field theories}},
  \href{http://dx.doi.org/10.1016/0550-3213(93)90548-4}{{\sf AMS/IP Stud. Adv.
  Math.} {\sf {1} }{\sf (1996) }{\sf 655--682}},
  \href{http://arxiv.org/abs/hep-th/9302103}{{\ttfamily arXiv:hep-th/9302103}}.

\bibitem{Bershadsky:1993cx}
M.~Bershadsky, S.~Cecotti, H.~Ooguri, and C.~Vafa, {\it {Kodaira-Spencer theory
  of gravity and exact results for quantum string amplitudes}},
  \href{http://dx.doi.org/10.1007/BF02099774}{{\sf Commun. Math. Phys.} {\sf
  {165} }{\sf (1994) }{\sf 311--428}},
  \href{http://arxiv.org/abs/hep-th/9309140}{{\ttfamily arXiv:hep-th/9309140}}.

\bibitem{Loran:2010bd}
F.~Loran, M.~Sheikh-Jabbari, and M.~Vincon, {\it {Beyond Logarithmic
  Corrections to Cardy Formula}},
  \href{http://dx.doi.org/10.1007/JHEP01(2011)110}{{\sf JHEP} {\sf {01} }{\sf
  (2011) }{\sf 110}}, \href{http://arxiv.org/abs/1010.3561}{{\ttfamily
  arXiv:1010.3561 [hep-th]}}.

\bibitem{Hartman:2014oaa}
T.~Hartman, C.~A. Keller, and B.~Stoica, {\it {Universal Spectrum of 2d
  Conformal Field Theory in the Large c Limit}},
  \href{http://dx.doi.org/10.1007/JHEP09(2014)118}{{\sf JHEP} {\sf {09} }{\sf
  (2014) }{\sf 118}}, \href{http://arxiv.org/abs/1405.5137}{{\ttfamily
  arXiv:1405.5137 [hep-th]}}.

\bibitem{Poghossian:2009mk}
R.~Poghossian, {\it {Recursion relations in CFT and N=2 SYM theory}},
  \href{http://dx.doi.org/10.1088/1126-6708/2009/12/038}{{\sf JHEP} {\sf {12}
  }{\sf (2009) }{\sf 038}}, \href{http://arxiv.org/abs/0909.3412}{{\ttfamily
  arXiv:0909.3412 [hep-th]}}.

\bibitem{Liu:2018iki}
C.~Liu and D.~A. Lowe, {\it {Notes on Scrambling in Conformal Field Theory}},
  \href{http://dx.doi.org/10.1103/PhysRevD.98.126013}{{\sf Phys. Rev. D} {\sf
  {98} }{\sf no.~12, }{\sf (2018) }{\sf 126013}},
  \href{http://arxiv.org/abs/1808.09886}{{\ttfamily arXiv:1808.09886
  [hep-th]}}.

\bibitem{Hampapura:2018otw}
H.~R. Hampapura, A.~Rolph, and B.~Stoica, {\it {Scrambling in Two-Dimensional
  Conformal Field Theories with Light and Smeared Operators}},
  \href{http://dx.doi.org/10.1103/PhysRevD.99.106010}{{\sf Phys. Rev. D} {\sf
  {99} }{\sf no.~10, }{\sf (2019) }{\sf 106010}},
  \href{http://arxiv.org/abs/1809.09651}{{\ttfamily arXiv:1809.09651
  [hep-th]}}.

\bibitem{Chang:2018nzm}
C.-M. Chang, D.~M. Ramirez, and M.~Rangamani, {\it {Spinning constraints on
  chaotic large $c$ CFTs}},
  \href{http://dx.doi.org/10.1007/JHEP03(2019)068}{{\sf JHEP} {\sf {03} }{\sf
  (2019) }{\sf 068}}, \href{http://arxiv.org/abs/1812.05585}{{\ttfamily
  arXiv:1812.05585 [hep-th]}}.

\bibitem{Kraus:2018zrn}
P.~Kraus, A.~Sivaramakrishnan, and R.~Snively, {\it {Late time Wilson lines}},
  \href{http://dx.doi.org/10.1007/JHEP04(2019)026}{{\sf JHEP} {\sf {04} }{\sf
  (2019) }{\sf 026}}, \href{http://arxiv.org/abs/1810.01439}{{\ttfamily
  arXiv:1810.01439 [hep-th]}}.

\bibitem{Kusuki:2018wpa}
Y.~Kusuki, {\it {Light Cone Bootstrap in General 2D CFTs and Entanglement from
  Light Cone Singularity}},
  \href{http://dx.doi.org/10.1007/JHEP01(2019)025}{{\sf JHEP} {\sf {01} }{\sf
  (2019) }{\sf 025}}, \href{http://arxiv.org/abs/1810.01335}{{\ttfamily
  arXiv:1810.01335 [hep-th]}}.

\bibitem{Wyllard:2009hg}
N.~Wyllard, {\it {A(N-1) conformal Toda field theory correlation functions from
  conformal N = 2 SU(N) quiver gauge theories}},
  \href{http://dx.doi.org/10.1088/1126-6708/2009/11/002}{{\sf JHEP} {\sf {11}
  }{\sf (2009) }{\sf 002}}, \href{http://arxiv.org/abs/0907.2189}{{\ttfamily
  arXiv:0907.2189 [hep-th]}}.

\bibitem{Beccaria:2015shq}
M.~Beccaria, A.~Fachechi, and G.~Macorini, {\it {Virasoro vacuum block at
  next-to-leading order in the heavy-light limit}},
  \href{http://dx.doi.org/10.1007/JHEP02(2016)072}{{\sf JHEP} {\sf {02} }{\sf
  (2016) }{\sf 072}}, \href{http://arxiv.org/abs/1511.05452}{{\ttfamily
  arXiv:1511.05452 [hep-th]}}.

\bibitem{Bombini:2018jrg}
A.~Bombini, S.~Giusto, and R.~Russo, {\it {A note on the Virasoro blocks at
  order $1/c$}},  \href{http://dx.doi.org/10.1140/epjc/s10052-018-6522-5}{{\sf
  Eur. Phys. J. C} {\sf {79} }{\sf no.~1, }{\sf (2019) }{\sf 3}},
  \href{http://arxiv.org/abs/1807.07886}{{\ttfamily arXiv:1807.07886
  [hep-th]}}.

\bibitem{Fitzpatrick:2016thx}
A.~L. Fitzpatrick and J.~Kaplan, {\it {A Quantum Correction To Chaos}},
  \href{http://dx.doi.org/10.1007/JHEP05(2016)070}{{\sf JHEP} {\sf {05} }{\sf
  (2016) }{\sf 070}}, \href{http://arxiv.org/abs/1601.06164}{{\ttfamily
  arXiv:1601.06164 [hep-th]}}.

\bibitem{stackexchange}
\url{https://mathematica.stackexchange.com/a/89682}.

\end{thebibliography}\endgroup
\bibliographystyle{bibstyle2017}

\end{document}